\documentclass[a4paper,fleqn]{cas-sc}

\usepackage[numbers,sort]{natbib} 
\usepackage[ruled, linesnumbered,lined,boxed]{algorithm2e}
\usepackage{graphicx} 
\usepackage{subfigure}
\usepackage{enumitem}

\setenumerate[1]{itemsep=0pt,partopsep=0pt,parsep=\parskip,topsep=5pt}
\setitemize[1]{itemsep=0pt,partopsep=0pt,parsep=\parskip,topsep=5pt}
\setdescription{itemsep=0pt,partopsep=0pt,parsep=\parskip,topsep=5pt}

\usepackage[defaultcolor=blue]{changes}  
\definechangesauthor[name={Author}, color=blue]{A} 

\def\tsc#1{\csdef{#1}{\textsc{\lowercase{#1}}\xspace}}
\tsc{WGM}
\tsc{QE}
\tsc{EP}
\tsc{PMS}
\tsc{BEC}
\tsc{DE}

\begin{document}
\let\WriteBookmarks\relax

\renewcommand{\topfraction}{0.8}
\renewcommand{\bottomfraction}{0.8}
\renewcommand{\textfraction}{0.2}
\renewcommand{\floatpagefraction}{0.8}
\let\printorcid\relax
\shorttitle\relax
\shortauthors{F.~Chen, Y.~Wang, C.~Lou, M.~Gao, and Q.~Xiong}


\title [mode = title]{SWGCN: Synergy Weighted Graph Convolutional Network for Multi-Behavior Recommendation}

\author[1]{Fangda Chen}
\ead{fangd.chen@gmail.com}

\affiliation[1]{
            organization={School of Big Data \& Software Engineering},
            addressline={Chongqing University}, 
            city={Chongqing},
            postcode={401331}, 
            country={China}}

\author[1]{Yueyang Wang}
\cormark[1]
\ead{yueyangw@cqu.edu.cn}
\cortext[1]{Corresponding author}

\author[1]{Chaoli Lou}
\ead{louchaoli@stu.cqu.edu.cn}

\author[1]{Min Gao}
\ead{gaomin@cqu.edu.cn}

\author[1]{Qingyu Xiong}
\ead{xiong03@cqu.edu.cn}

\begin{abstract}
Multi-behavior recommendation paradigms have emerged to capture diverse user activities, forecasting primary conversions (e.g., purchases) by leveraging secondary signals like browsing history. However, current graph-based methods often overlook cross-behavioral synergistic signals and fine-grained intensity of individual actions. Motivated by the need to overcome these shortcomings, we introduce \textbf{S}ynergy \textbf{W}eighted \textbf{G}raph \textbf{C}onvolutional \textbf{N}etwork (\textbf{SWGCN}). SWGCN introduces two novel components: a Target Preference Weigher, which adaptively assigns weights to user-item interactions within each behavior, and a Synergy Alignment Task, which guides its training by leveraging an Auxiliary Preference Valuator. This task prioritizes interactions from synergistic signals that more accurately reflect user preferences. The performance of our model is rigorously evaluated through comprehensive tests on three open-source datasets, specifically Taobao, IJCAI, and Beibei. On the Taobao dataset, SWGCN yields relative gains of 112.49\% and 156.36\% in terms of Hit Ratio (HR) and Normalized Discounted Cumulative Gain (NDCG), respectively. It also yields consistent gains on IJCAI and Beibei, confirming its robustness and generalizability across various datasets. Our implementation is open-sourced and can be accessed via \href{https://github.com/FangdChen/SWGCN}{https://github.com/FangdChen/SWGCN}.

\end{abstract}



\begin{keywords}
 Multi-Behavior Recommendation \sep Graph Neural Network 
\end{keywords}

\maketitle

\section{Introduction}

Recommendation systems have become an integral component of the digital landscape, with pervasive implementations in online retail giants such as Amazon and Alibaba’s Taobao, along with popular video streaming platforms like Netflix and YouTube. These systems effectively mitigate information overload, enhance user engagement, and boost platform profitability. Over the last ten years, recommendation systems have garnered significant interest among scholars. Collaborative Filtering (CF)~\cite{sedhain2015autorec,wu2016collaborative,yang2021hyper} has become a leading and successful approach in this field. The primary concept of the CF technique is to convert user preferences and item characteristics into low-dimensional vectors using historical data. Traditional CF methods primarily focus on single target behavior interactions, typically linked to platform monetization, such as purchases. Representative approaches in this category include Neural Collaborative Filtering (NCF)~\cite{ncf}.

Recently, multi-behavior recommendation has emerged as a key research domain~\cite{xia2021multi, xia2021graph, xia2020multiplex}. This approach utilizes various types of behaviors for recommendation tasks, with researchers observing that the integration of diverse behavior interactions provides a broader insight into user preferences. A primary motivation is that recommendation systems often suffer from severe data sparsity when relying exclusively on target interactions~\cite{mbgcn}. Additionally, different behavior types encapsulate distinct semantics and reflect richer user preferences~\cite{yang2021hyper, xia2021multi, xia2021graph}. Early research efforts have aimed to extend Matrix Factorization (MF)~\cite{koren2009matrix} techniques to enable collaborative learning across multiple behaviors. For instance, Collective Matrix Factorization (CMF)~\cite{singh2008relational} performs joint factorization on multiple relational matrices. Additional studies~\cite{loni2016bayesian,qiu2018bprh} address the task through learning methodologies that model the influence of multiple behavior types with adjustable parameters. Multi-Feedback Bayesian Personalized Ranking (MF-BPR)~\cite{loni2016bayesian} extends Bayesian Personalized Ranking (BPR)~\cite{bprmf} and maps different feedback channels to distinct hierarchical levels. Some other works, such as NMTR~\cite{nmtr}, attempt to model the sequential dependency across different behavior types by learning transformation functions. However, these approaches either adopt rigid assumptions about behavior dependencies or treat all behaviors as equally important, which limits their flexibility in modeling real-world user behavior dynamics.

Advancements in Graph Neural Network (GNN) architectures have encouraged scholars to employ these models for representing user-item interactions within recommendation frameworks~\cite{nmtr, ghcf, xia2021graph}. Within these frameworks, users and items function as vertices, with diverse interaction types being captured through distinct classes of edges. The primary objective of such GNN-powered recommenders involves deriving latent embeddings for both users and products by exploring the underlying two-mode interaction network. For instance, Neural Graph Collaborative Filtering (NGCF)~\cite{ngcf} enhances the characterization of user and product embeddings by leveraging Graph Convolutional Networks (GCNs)~\cite{gcn} over bipartite interaction networks. Subsequently, LightGCN~\cite{lightgcn} refines the NGCF approach to further enhance interaction modeling capabilities. Drawing inspiration from previous multi-behavior CF methods, some researchers have integrated GNNs into multi-behavior recommendation frameworks. Multi-Behavior Graph Convolutional Network (MBGCN)~\cite{mbgcn} assigns distinct weight parameters to edges corresponding to different behaviors. Additionally, Graph Heterogeneous Collaborative Filtering (GHCF)~\cite{ghcf} integrates user, item, and multi-behavior latent features within a single, cohesive GNN architecture for joint optimization. Despite their progress, these models generally apply fixed weighting schemes for different behaviors that fail to reflect the actual synergy of behaviors per interaction. Moreover, they do not explicitly quantify or utilize the co-occurrence signals between auxiliary and target behaviors in a fine-grained and behavior-specific manner.

\begin{figure}[pos=!t]
    \centering
	\includegraphics[width=0.5\linewidth]{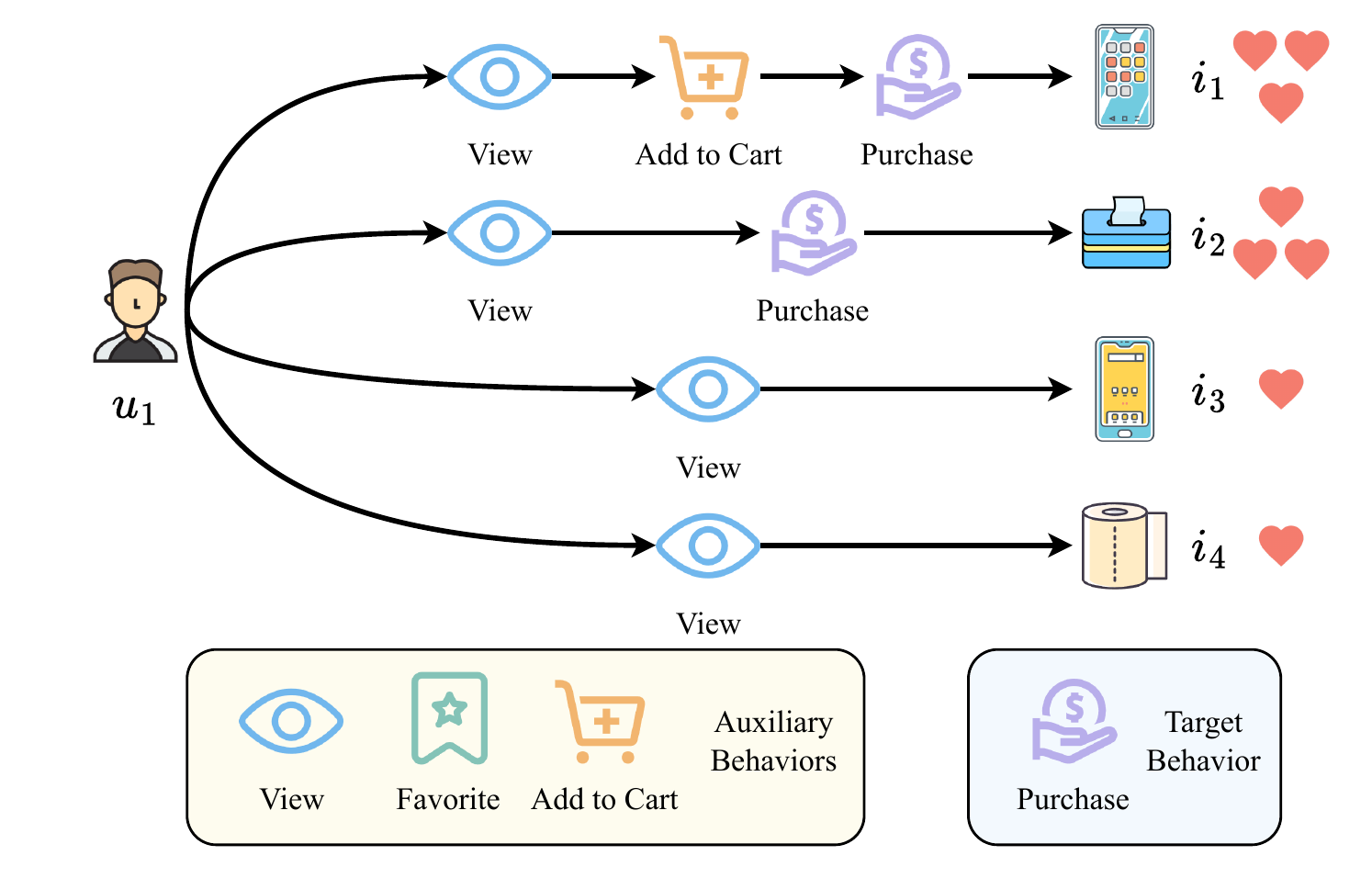}
    \caption{A case of the synergistic relationships between the target behavior and auxiliary behaviors.}
    \label{fig: synergistic signals}
    \vspace{-0.5cm}
\end{figure}

The motivation for this work stems from key limitations observed in real-world multi-behavior settings, particularly in the ability to identify fine-grained interaction signals and to effectively capture the synergy between behaviors, as illustrated in~\ref{fig: synergistic signals}. Addressing this issue requires overcoming two primary limitations: (1) \textbf{The lack of an approach to identify the diversity impacts of interactions within a single behavior.} Most representation learning under a single behavior assumes a uniform impact of all interactions and the corresponding items. Conversely, items exert varying influences on a single behavior, particularly for auxiliary behaviors. For instance, in viewing behavior, viewed and purchased items more accurately epitomize user preferences than those viewed alone. As shown in~\ref{fig: synergistic signals}, for the four viewed items, user $u_1$ obviously prefers items $i_1$ and $i_2$ for which the purchase behavior occurs. Consequently, an interaction modeling approach is essential to delineate the differential influence strengths of various items under a single behavior. (2) \textbf{The absence of a method for modeling synergistic signals between multiple behaviors.} Existing approaches assume fixed behavior patterns across all item interactions, leading to coarse-grained modeling of behavioral synergistic information. For example, methods that learn a fixed synergistic pattern between viewing and purchasing behaviors assume that all viewed items are subsequently more prone to be purchased. However, user behavior patterns are complicated across different items. ~\ref{fig: synergistic signals} illustrates the disparity in user $u_1$’s activity records concerning the smartphone $i_1$ versus the paper towel $i_2$. Users may exhibit a view-then-purchase pattern for everyday necessities like paper towels, whereas a cart-then-purchase pattern may be more prevalent for premium durables such as smartphones. The synergistic signals between behaviors are not uniform across different items. Therefore, it is imperative to fine-grainly model behavioral synergistic signals that discern the various synergistic scenarios for different items.

To overcome the restricted modeling power of existing GNN-centric frameworks for multi-behavior tasks, particularly their insufficient modeling of interaction intensity within a single behavior and synergy across behaviors, we introduce a novel architecture: the \textbf{S}ynergy \textbf{W}eighted \textbf{G}raph \textbf{C}onvolutional \textbf{N}etwork (\textbf{SWGCN}). Specifically, SWGCN introduces two core components that jointly model fine-grained intra-behavior interaction strengths and inter-behavior synergy:
(1) \textbf{Target Preference Weigher.} Based on the correspondence between user-side and item-side feature vectors, the module calculates edge significance for a distinct interaction mode within the bipartite framework. Such magnitudes signify connection strengths, empowering the architecture to isolate the diverse contributions of distinct items toward user-side profiles for any given activity type. This fine-grained modeling enables adaptive learning beyond fixed interaction types or one-hot behavior encodings.
(2) \textbf{Synergy Alignment Task.} We introduce a training objective that computes the alignment loss between interaction strengths under target behavior (from Target Preference Weigher) and auxiliary behavior (from Auxiliary Preference Valuator). This task effectively reduces the impact of items involved only in auxiliary behaviors and increases the weights of items appearing in both target and auxiliary behaviors, thereby explicitly modeling cross-behavior synergy.

Compared to prior approaches, our model differs in several key aspects. MBGCN~\cite{mbgcn} and GHCF~\cite{ghcf} rely on fixed item weighting mechanisms that do not fully capture item-wise interaction importance. END4Rec~\cite{end4rec} and MB-CGCN~\cite{mbcgcn} model different behaviors with separate embeddings or sequential transfer paths, yet they lack a unified mechanism to learn synergistic signals between behaviors. These limitations hinder their ability to accurately reflect user intent when interactions are ambiguous or sparse.

\vspace{1em}
\noindent\textbf{This study contributes to the field in the following ways:}

\begin{itemize}[leftmargin=*] 
\item We introduce \textbf{Target Preference Weigher}, a component designed to quantify the strength of intra-behavioral engagement by assigning proximity-driven weights to links connecting user-product pairs, thereby facilitating more nuanced capture of latent inclinations within each behavior type.

\item We design the \textbf{Synergy Alignment Task}, a contrastive-style auxiliary objective that aligns interaction weights across behaviors. This facilitates the identification of synergistic signals between behaviors, providing a principled optimization perspective that is missing in previous modular or sequential models.

\item We present an adaptive architecture titled \textbf{SWGCN}, which addresses the complexities of multi-action data by synchronizing intra-behavioral diversity with cross-behavioral synergy. Empirical evidence derived from three large-scale real-world scenarios confirms the superiority of SWGCN over existing strong benchmarks. On the Taobao dataset, our model achieves a remarkable performance leap, increasing HR by 112.49\% and NDCG by 156.36\%.
\end{itemize}

\section{Related Work}

\subsection{Multi-Behavior Recommendation}

To enhance prediction accuracy, modern recommendation engines harness a wide array of user-item interaction modes rather than relying on a single signal. Early research extended MF~\cite{koren2009matrix} to facilitate joint learning across multiple behaviors~\cite{singh2008relational,krohn2012multi,tang2016empirical}. Specifically, CMF~\cite{singh2008relational} performs joint factorization on multiple relational matrices to capture inter-behavior correlations. Additionally, some researchers have approached the recommendation challenge through various learning methodologies~\cite{loni2016bayesian,qiu2018bprh}, suggesting modeling of diverse influences through learnable parameters. MF-BPR~\cite{loni2016bayesian} enhances BPR~\cite{bprmf} by associating different behaviors with distinct levels, thereby illustrating the contribution of each behavior type. Other studies have investigated the assignment of various learnable weights to specific behavior types during the training of MF~\cite{qiu2018bprh}.

Lately, GNN-oriented recommendation frameworks have witnessed a remarkable evolution~\cite{gu2022self, wang2020beyond, kipi}. These methodologies conceptualize users and items as nodes, with multi-behavior interactions representing different edge types connecting these nodes. MBGCN~\cite{mbgcn} constructs a unified heterogeneous graph that incorporates user multi-feedback, assigning distinct weight parameters to edges and utilizing the robust learning capabilities of GCNs to embed users and items. GHCF~\cite{ghcf} employs a link-sensitive GCN message-passing module, which feeds the fused latent descriptors of vertices and edges into a joint optimization scheme. Multiplex Graph Neural Network (MGNN)~\cite{zhang2020multiplex} proposes learning unique embeddings for each user or item corresponding to each behavior, thereby facilitating targeted recommendations for specific behaviors. MB-CGCN~\cite{mbcgcn} and END4Rec~\cite{end4rec} explicitly model cascading dependencies to learn sequential correlations among multiple types of user behaviors. In BIPN~\cite{bipn}, secondary interaction signals are reserved uniquely for calibrating the underlying architecture, a tactic designed to dampen the interference inherent in such data. Notwithstanding the success of these graph-based paradigms, they often rely on rigid behavior-level fusion or fixed edge importance, limiting their capacity to fully exploit the intra-behavior diversity and cross-behavior synergy that are critical for fine-grained user preference modeling.

\subsection{Graph Neural Networks}

The unparalleled efficacy of GNNs in distillation of latent features has propelled them to the forefront of academic inquiry throughout the contemporary era. Initial research efforts extended convolutional operations to graph data, resulting in the GCN~\cite{gcn}, which aggregates local node information and neighborhood data to update node features. Later, the integration of attention-based weightings into neighborhood message-passing fostered the emergence of the Graph ATtention Network (GAT)~\cite{gat}. For the purpose of parsing the complex connectivity patterns inherent in graph-structured domains, Graph Isomorphism Network (GIN)~\cite{xu2018powerful} and the Topology Adaptive Graph Convolutional Network (TAG)~\cite{du2017topology} are proposed, both grounded in geometric and topological principles. Inspired by the powerful representation learning ability of GNNs, some researchers have attempted to apply GNNs to recommendation systems. For instance, DGS-MGNN~\cite{dgsmgnn} introduces a multi-channel graph neural network to enhance the modeling efficacy of session-based recommendation.

\section{Preliminary}

\subsection{Problem Statement} 
Suppose there are $R$ unique action classes; we assign the $R$-th index to the target behavior (exemplified by purchasing), whereas the residual categories are classified as auxiliary behaviors. By leveraging historical interactions, we construct a series of interaction matrices corresponding to the $R$ types of behaviors, denoted as $\{\mathcal{A}_1,\ldots,\mathcal{A}_R\}$, in which every $\mathcal{A}_r \in \mathbb{R}^{N_u \times N_i}$ is a relational matrix, with the population sizes of users and items denoted by $N_u$ and $N_i$. For any $(u, i)$ pair, the value of $a_{u,i;r} \in \mathcal{A}_r$ becomes 1 if a record exists for user $u$ and item $i$ regarding the $r$-th interaction category. Conversely, the element is defined as 0. Given the interaction matrices, the target of recommendation systems is to forecast future interactions of the target behavior by utilizing historical interactions of auxiliary and target behaviors. Accordingly, we reformulate the problem specified as:
\begin{itemize}[leftmargin=*] 
    \item \textbf{Input:} The sequence of user-item adjacency grids $\{\mathcal{A}_r\}_{r=1}^R$.
    \item \textbf{Output:} The predicted propensity score for the $R$-th interaction between user $u$ and item $i$.
\end{itemize}

\subsection{Graph Convolutional Network}
This part provides a concise summary of the graph encoding mechanism employed by the vanilla GCN~\cite{gcn}. Let $\mathcal{G}$ be a graph with $N$ individual points. The relationship between these points is encoded in $\mathcal{A} \in \mathbb{R}^{N \times N}$, which defines the link topology. Additionally, $\mathcal{E} \in \mathbb{R}^{N \times d}$ is the comprehensive matrix containing $d$-dimensional embeddings for all respective points, where $d$ signifies the embedding dimension. The GCN uses the following formula for propagation:
\begin{equation}
\mathcal{E}^{(l)} = \sigma(\mathcal{D}^{-1/2} \mathcal{A} \mathcal{D}^{-1/2} \mathcal{E}^{(l-1)}\mathcal{W}^{(l-1)}).
\end{equation}

In this formulation, $l \in \{1,\ldots,L\}$ represents the layer index; the degree matrix $\mathcal{D}$ is computed from $\mathcal{A}$ such that its diagonal elements $D_{i,i}$ equal $\sum_{j=1}^N \mathcal{A}_{i,j}$. The trainable weights for the current step are contained in $\mathcal{W}^{(l-1)} \in \mathbb{R}^{d_{(l-1)} \times d_{(l)}}$, while $\sigma(\cdot)$ signifies a non-linear transformation, for instance, the $ReLU(\cdot)$ operation. Furthermore, $\mathcal{E}^{(l)} \in \mathbb{R}^{N \times d_{(l)}} $ captures the nodal latent descriptors at depth $l$, where $d_{(l)}$ indicates the corresponding hidden size.

\section{Methodology}

\begin{figure*}
\centering
\includegraphics[width=\linewidth]{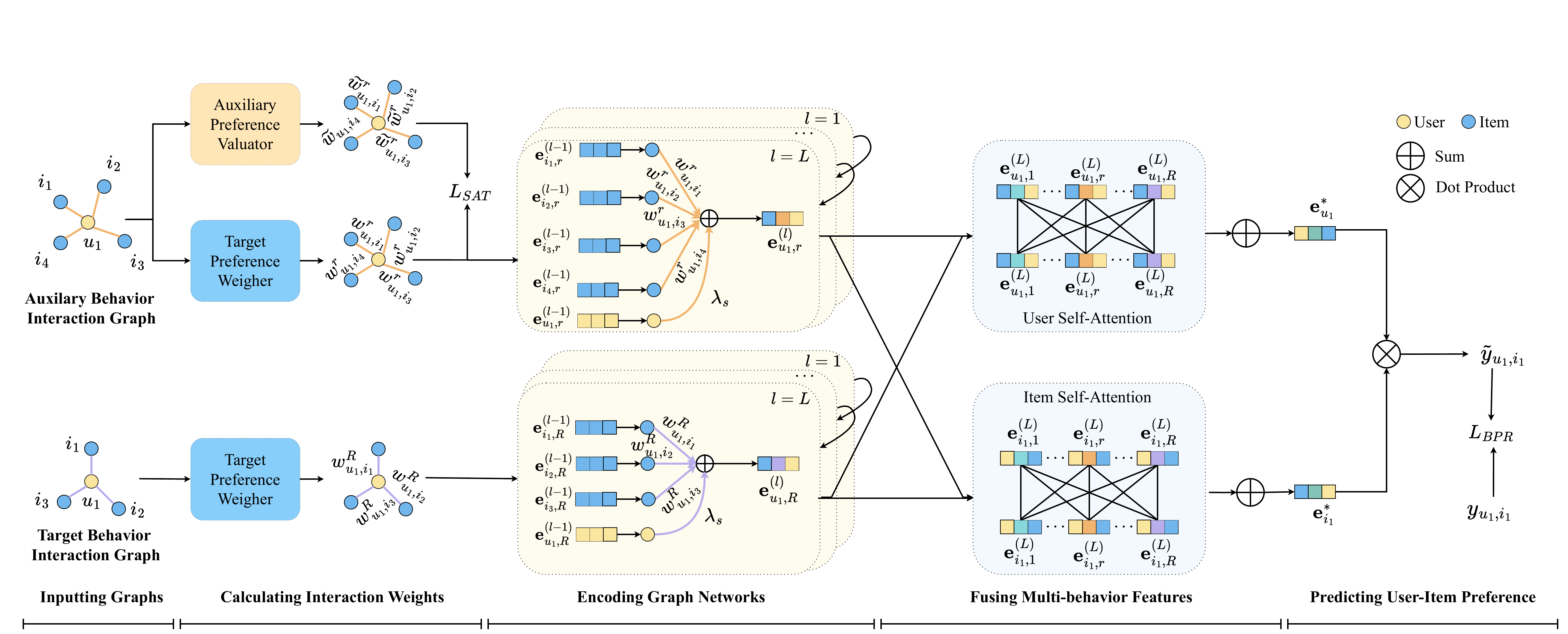}
\caption{The overall framework of SWGCN. 
(1) Input the user-item bipartite graphs of auxiliary and target behaviors. 
(2) Calculate interaction weights for each behavior using the Target Preference Weigher. The edge weights $w_{u_1,i_1;r},w_{u_1,i_2;r},w_{u_1,i_3;r},w_{u_1,i_4;r}$  represent the interaction strengths between user $u_1$ and items $i_1,i_2,i_3,i_4$ under $r$-th behavior, respectively. Likewise, $w_{u_1,i_1;R}, w_{u_1,i_2;R},$ and $w_{u_1,i_3;R}$ serve to quantify the engagement intensities associated with the $R$-th behavior. The auxiliary behavior graph undergoes the Auxiliary Preference Valuator to estimate interaction weights based on auxiliary behavior preferences, facilitating the computation of the Synergy Alignment Task loss, $L_{SAT}$.
(3) Encode node features in the user-item bipartite graphs with interaction weights via graph neural networks. In each layer, nodes update their features by aggregating neighboring node features based on the interaction weights and self-loop weight $\lambda_s$. For instance, under the $r$-th behavior, user $u_1$ updates its $l$-th layer node feature $\mathbf{e}_{u_1,r}^{(l)}$ by aggregating the previous layer features of neighboring items, $\mathbf{e}_{i_1,r}^{(l-1)}, \mathbf {e}_{i_2,r}^{(l-1)},\mathbf{e}_{i_3,r}^{(l-1)},\mathbf{e}_{i_4,r}^{(l-1)}$, and its own upper layer features $\mathbf{e}_{u_1,r}^{(l-1)}$. A parallel process is executed for the target behavior. The $L$-th layer features are used as the encoded features under behaviors $1,\ldots,r,\ldots,R$ such as $\mathbf{e}_{u_1,1}^{(L)},\ldots, \mathbf{e}_{u_1,r}^{(L)}, ..., \mathbf{e}_{u_1,R}^{(L)}$. 
(4) Fuse multi-behavior user and item features through a self-attention mechanism, and summate to obtain the final feature representations, e.g., $\mathbf{e}_{u_1}^*, \mathbf{e}_{i_1}^*$. 
(5) Predict the preference scores through the dot product between the ultimate user and item embeddings, e.g., $\widetilde{y}_{u_1,i_1}$, denoting the preference score of user $u_1$ for item $i_1$. Then the BPR loss $L_{BPR}$ is computed using the true preference, e.g., $y_{u_1,i_1} \in \{0,1\}$, indicating the occurrence of interaction under the target behavior.}

\label{overall framework}
\vspace{-0.5cm}
\end{figure*}

We delineate the structural design of SWGCN in this part, which is visualized in~\ref{overall framework}. The proposed methodology is partitioned into five sequential phases. (1) Inputting Graphs: Input bipartite connectivity networks derived from the collective history of user-item engagements. (2) Calculating Interaction Weights: Calculate the interaction weights of the user-item interaction bipartite graphs using the Target Preference Weigher. (3) Encoding Graph Networks: Encode the weighted user-item interaction bipartite graphs using graph neural networks. (4) Fusing Multi-behavior Features: Fuse multiple user or item features across various behaviors by a self-attention mechanism. (5) Predicting User-Item Preference: Estimate the interaction likelihood by evaluating the correspondence between user and item feature descriptors.

\subsection{Inputting Graphs}
When constructing the bipartite topology, it is imperative to assign starting attribute descriptors for the respective nodes before any message-passing occurs. In contrast to traditional multi-behavior paradigms that rely on behavior-invariant representations~\cite{mbgcn}, SWGCN treats the latent descriptors of users and products as decoupled across different interaction types. Specifically, we employ a stochastic assignment process to establish the starting feature states for every nodal entity within each individual behavioral domain. These initial embeddings are designated as $0$-th layer embeddings, denoted as $ \mathcal{E}_{r}^{(0)} \in \mathbb{R}^{(N_u+N_i) \times d}, r \in \{1,...R\} $, where $d$ signifies the embedding dimension.

\subsection{Calculating Interaction Weights}

\subsubsection{Target Preference Weigher}
Within user-item bipartite graph structures, the Target Preference Weigher assigns interaction weights to each graph. The scalar $w_{u,i;r} \in \mathcal{W}_r$ serves to quantify the connectivity magnitude between user $u$ and item $i$ specific to the $r$-th action category, which is formulated as follows:
\begin{equation}
    w_{u,i;r} = \frac{exp(ELU(|\mathbf{e}_{u,r}^{(0)}-\mathbf{e}_{i,r}^{(0)}|\mathbf{\beta}_r^T))} {\sum_{j\in \mathcal{N}_r(u)}{exp(ELU(|\mathbf{e}_{u,r}^{(0)}-\mathbf{e}_{j,r}^{(0)}|\mathbf{\beta}_r^T))}}, \\
\end{equation}
where the starting embeddings of user $u$ and item $i$ regarding the $r$-th action category are denoted by $\mathbf{e}_{u,r}^{(0)}, \mathbf{e}_{i,r}^{(0)} \in \mathcal{E}_{r}^{(0)}$; $\beta_r \in \mathbb{R}^{1 \times d}$ denotes mapping weight for the $r$-th behavior; $\mathcal{N}_r(u)$ indicates the collection of adjacent nodes for the user $u$ under the $r$-th behavior, and $ELU(\cdot)$ indicates the activation function. Specifically, the Target Preference Weigher facilitates the concurrent derivation of relational intensities $\mathcal{W}_r$ throughout the entire bipartite topology $\mathcal{A}_r$.

\subsubsection{Auxiliary Preference Valuator}
To leverage the synergistic signals between the target and auxiliary behaviors, we propose integrating preferences from both behavior types to refine the learning of interaction weights. The target behavior preferences are derived from the prediction task associated with target behavior interactions, while the auxiliary behavior preferences are calculated using the following equation:
\begin{equation}
    \widetilde{w}_{u,i;r} =\left\|\mathbf{e}_{u,r}^{(0)} - \mathbf{e}_{i,r}^{(0)}\right\|_2^2, 
\end{equation}
where $\widetilde{w}_{u,i;r} \in \widetilde{\mathcal{W}}_r$ denotes the auxiliary behavioral preference score and $\widetilde{\mathcal{W}}_r \in \mathbb{R}^{N_u \times N_i}$. Note that these behaviors contain only auxiliary behaviors, i.e., $r \in \{1,\ldots, R-1\}$.

\subsection{Encoding Gragh Networks}
Given the user-item bipartite graphs characterized by interaction weights, a GNN is employed to encode the graph, facilitating the extraction of features for both users and items across diverse behaviors. To effectively capture information from multiple hops, the aggregation of neighboring node information is conducted layer by layer. Drawing inspiration from LightGCN~\cite{lightgcn}, the aggregation process utilizes simple weighting and integrates neighbor information, intentionally excluding feature transformations and nonlinear activation functions. However, empirical investigations have indicated the necessity of retaining self-loops. To be precise, the primary nodal descriptors remain intact, while the scalar $\lambda_s$ serves to modulate the magnitude of self-referential connections. An extensive experimental analysis is detailed in Section~\ref{self_loop_weight_exp}. The corresponding aggregation rule is defined as follows:
\begin{gather}
    \mathbf{e}_{u,r}^{(l)} = (\frac{\lambda_s \mathbf{e}_{u,r}^{(l-1)}}{\sqrt{d_{u,r}^2}} + \sum_{j \in \mathcal{N}_{r}(u) }{\frac{w_{u,j;r} \mathbf{e}_{j, r}^{(l-1)}}{\sqrt{d_{u,r}d_{j,r}}} })) ,  \\ 
    \mathbf{e}_{i,r}^{(l)} = (\frac{\lambda_s \mathbf{e}_{i,r}^{(l-1)}}{\sqrt{d_{i,r}^2}} + \sum_{j \in \mathcal{N}_{r}(i) }{\frac{w_{j,i;r} \mathbf{e}_{j, r}^{(l-1)}}{\sqrt{d_{j,r}d_{i,r}}} })), 
\end{gather}
where $\mathbf{e}_{u,r}^{(l)}$ and $\mathbf{e}_{i,r}^{(l)}$ signify the d-dimensional features of user $u$ plus item $i$ for the $r$-th connection type in the $l$-th GNN propagation step. The embedding matrix is defined as $\mathcal{E}_r^{(l)} \in \mathbb{R}^{(N_u+N_i) \times d}$, where $l$ iterates from $1$ to $L$. Furthermore, $\mathcal{N}_{r}(\cdot)$ denotes the group of reachable vertices for a specific user or product in the context of behavior $r$, while the terms $d_{u,r}$, $d_{i,r}$, and $d_{j,r}$ refer to the respective degrees of nodes $u$, $i$, and $j$. Consequently, the matrix formulation describing the propagation process at step $l$ is expressed as:
\begin{gather}
     \widetilde{\mathcal{A}}_r = \begin{pmatrix} \lambda_s \mathcal{I}_1 & \mathcal{W}_r \odot \mathcal{A}_r \\ \mathcal{W}_r^T \odot \mathcal{A}_r^T & \lambda_s \mathcal{I}_2 \end{pmatrix}, \\
     \mathcal{E}_{r}^{(l)} = \widetilde{\mathcal{D}}_r^{-1/2} \widetilde{\mathcal{A}}_r \widetilde{\mathcal{D}}_r^{-1/2} \mathcal{E}_{r}^{(l-1)},
\end{gather}
where $\widetilde{\mathcal{A}}_r \in \mathbb{R}^{(N_u+N_i) \times (N_u+N_i)}$ denotes the augmented adjacency matrix with self-loops. Here, $\mathcal{I}_1 \in \mathbb{R}^{N_u \times N_u}$ and $\mathbf{I}_2 \in \mathbb{R}^{N_i \times N_i}$ correspond to identity matrices for the user and item blocks, respectively. The symbol $\odot$ represents the Hadamard product, while $\widetilde{\mathcal{D}}_r$ corresponds to the diagonal degree matrix obtained from $\widetilde{\mathcal{A}}_r$.

\subsection{Fusing Multi-behavior Features}
We leverage a self-attentive strategy to fuse latent factors, thereby consolidating information regarding users and items from diverse interaction types. Furthermore, residual connections are added to boost overall generalization capabilities. The formula for the self-attention integration is delineated below:
\begin{gather}
    \mathcal{Q}_{r} = \mathcal{E}_r^{(L)} \mathcal{W}_q, \\
    \mathcal{K}_{r} = \mathcal{E}_r^{(L)} \mathcal{W}_k, \\
    \mathcal{V}_{r} = \mathcal{E}_r^{(L)} \mathcal{W}_v, \\
    \widetilde{\mathcal{E}}_r = \mathcal{E}_r^{(L)} + \sum_{r^\prime}{Softmax\left(\frac{\mathcal{Q}_{r}\mathcal{K}_{r^\prime}^T}{\sqrt{d}}\right)\mathcal{V}_{r^\prime}},
\end{gather}
where the $r^\prime$-th behavior means other behaviors including the $r$-th behavior; $\mathcal{W}_q, \mathcal{W}_k, \mathcal{W}_v \in  \mathbb{R}^{d \times d}$ represent the weight parameters; and $Softmax(\cdot)$ denotes the softmax function. Subsequently, a simple union function $g(\cdot)$ is utilized to union embeddings across different behaviors, resulting in the ultimate embedding $\mathcal{E}^{*}$ for users and items, as presented:
\begin{equation}
    \mathcal{E}^*=g(\widetilde{\mathcal{E}}_1,...,\widetilde{\mathcal{E}}_R).
\end{equation}

A range of aggregation functions can be used, including $sum(\cdot),$ $mean(\cdot), concat(\cdot)$,  among others, with the current study opting for $sum(\cdot)$.

\subsection{Predicting User-Item Preference}
Thereafter, the unified embedding matrix is divided to obtain specific representations for users, $\mathcal{E}_u^*$, and items, $\mathcal{E}_i^*$:
\begin{equation}
    \mathcal{E}_u^*, \mathcal{E}_i^*=Split(\mathcal{E}^*),
    \label{eq: split embeddings}
\end{equation}
where $Split(\cdot)$ represents the separation operator, yielding the matrices $\mathcal{E}_u^* \in \mathbb{R}^{N_u \times d}$ and $\mathcal{E}_i^* \in \mathbb{R}^{N_i \times d}$. This step generates the ultimate latent vectors $\mathbf{e}_u^*$ and $\mathbf{e}_i^*$ specific to user $u$ and item $i$. Finally, the estimated value $\widetilde{y}_{u,i}$, indicating the probability of adoption under the target behavior, is determined via the dot product:
\begin{equation}
    \widetilde{y}_{u,i} = {\mathbf{e}_u^*}{\mathbf{e}_i^*}^T.
    \label{eq: model prediction}
\end{equation}

\subsection{Model Training}
\subsubsection{Synergy Alignment Task} 
To exploit the synergistic signals between auxiliary and target behaviors for learning interaction weights, we introduce a novel training task termed the Synergy Alignment Task. This task introduces the Auxiliary Preference Valuator, which estimates interaction weights related to auxiliary behavioral preferences. Combined with the Target Preference Weigher, which infers interaction weights based on target behavioral preferences, this task calculates the preference alignment loss for both auxiliary and target behaviors. The interactions encompassing both auxiliary and target behaviors initially exhibit a lower loss. Conversely, interactions that involve solely auxiliary behaviors experience a gradual decrease in loss throughout the training iterations, guided by their preference scores from target behaviors. Finally, this task ultimately facilitates a progressive convergence. The exact formula for calculation is described here:
\begin{equation}
    L_{SAT}= \frac{1}{R}\sum_{r}{\frac{1}{N_r}\sum_{N_r}{(\mathcal{W}_r-\widetilde{\mathcal{W}}_r)}+ \frac{\gamma_1}{R}\sum_r{\left\|\widetilde{\mathcal{W}}_r\right\|_2^2}},
\end{equation}
in which $N_r$ signifies a total sum of interactions associated with the $r$-th behavior; $\gamma_1$ denotes $L_2$ regularization coefficient employed to mitigate overfitting.

\subsubsection{User-Item Prediction Task} 

Concurrently, we utilize the BPR criterion~\cite{bprmf} as the target interaction prediction loss, which is widely employed for optimizing pairwise training loss within the domain~\cite{mbgcn}. The BPR methodology emphasizes that items with observed interactions more accurately reflect user preferences and should consequently be assigned higher scores than items lacking observed interactions. Consequently, the objective function is expressed as:
\begin{equation}
    L_{BPR}=\sum_{(u,i,j) \in O} -\ln{\sigma(\widetilde{y}_{u,i}-\widetilde{y}_{u,j})}+\gamma_2 \left\|\Theta\right\|_2^2,
\end{equation} 
where $\sigma(\cdot)$ refers to the logistic sigmoid function, $\Theta$ denotes the complete set of learnable weights, and $\gamma_2$ acts as the weight decay hyperparameter for $L_2$ regularization to facilitate model generalization. Furthermore, $O=\{ (u,i,j) \mid (u,i) \in \chi^+, (u,j)\in \chi^- \} $ defines a grouping of pairwise triplets during the training stage, where $\chi^+$ and $\chi^-$ correspond to the pools of positive and negative feedback, respectively.

\subsubsection{Joint Objective Function} 
During the model training process, we simultaneously train the aforementioned two tasks, utilizing the hyperparameter $\lambda_a$ to regulate the weight of the Synergy Alignment Task, calculated as follows:
\begin{equation}
    Loss= \lambda_a \cdot L_{SAT} + (1- \lambda_a) \cdot L_{BPR}.
\end{equation}

\subsubsection{Message Dropout}
We implement a random dropout strategy to address the issue of model overfitting. For GNNs, node dropout and message dropout are commonly employed strategies to prevent overfitting during model training~\cite{ngcf, mbgcn}. However, in the proposed SWGCN, we argue that the explicit application of node dropout is redundant. This is because the Target Preference Weigher adaptively assigns varying interaction weights based on preference relevance; specifically, it assigns significantly lower weights to irrelevant or noisy nodes. This mechanism effectively functions as an implicit dropout strategy by attenuating the propagation of non-informative signals. Consequently, we strictly apply message dropout to the $l$-th user-item embedding matrices $\mathcal{E}_r^{(l)}$ (where $r \in \{1, \ldots, R\}$) with a dropout ratio of $p_{message}$.

\begin{algorithm}
\caption{Synergy Weighted Graph Convolutional Network} \label{algorithm}
\SetKwInOut{Input}{Input}\SetKwInOut{Output}{Output}
\SetKwFunction{TargetPreferenceWeigher}{TargetPreferenceWeigher}
\Input{$\{\mathcal{A}_1, \ldots, \mathcal{A}_R\}$, the collection of adjacency matrices representing diverse behaviors with $\mathcal{A}_r \in \mathbb{R}^{N_u \times N_i}$}
\Output{$\widetilde{y}_{u,i} \in [0,1]$, the estimated propensity for user $u$ to select item $i$ regarding the specific target behavior}
Initialize model parameters, $\mathcal{E}_r^{(0)} \in \mathbb{R}^{(N_u+N_i) \times d}, \beta_r \in \mathbb{R}^{1 \times d}, r=1,\ldots,R$, and ${\mathcal{W}_q, \mathcal{W}_k, \mathcal{W}_v} \in \mathbb{R}^{d \times d} $\;
\For{behavior $r \leftarrow 1$ \KwTo $R$}{
    \lForEach{edge $<u,i>$ of the matrix $\mathcal{A}_r$}{\\
    \quad Select $\mathbf{e}_{u,r}^{(0)},\mathbf{e}_{i,r}^{(0)}$ from $\mathcal{E}_r^{(0)}$; \\
    \quad $w_{u,i;r} \leftarrow$ \TargetPreferenceWeigher{$\mathbf{e}_{u,r}^{(0)},\mathbf{e}_{i,r}^{(0)}$}, $w_{u,i;r} \in \mathcal{W}_r$}
    $\widetilde{\mathcal{A}}_r \leftarrow \begin{pmatrix} \lambda_s \mathcal{I}_1 & \mathcal{W}_r \odot \mathcal{A}_r \\ \mathcal{W}_r^T \odot \mathcal{A}_r^T & \lambda_s \mathcal{I}_2 \end{pmatrix}$\;
}
\For{layer $l \leftarrow 1$ \KwTo $L$}{
    Get degree matrix $\widetilde{\mathcal{D}}_r$ from $\widetilde{\mathcal{A}}_r$\;
    $\mathcal{E}_{r}^{(l)} \leftarrow \widetilde{\mathcal{D}}_r^{-1/2} \widetilde{\mathcal{A}}_r \widetilde{\mathcal{D}}_r^{-1/2} \mathcal{E}_{r}^{(l-1)}$\;
}
\For{behavior $r \leftarrow 1$ \KwTo $R$}{
    $\mathcal{Q}_{r} \leftarrow \mathcal{E}_r^{(L)} \mathcal{W}_q$\;
    $\mathcal{K}_{r} \leftarrow \mathcal{E}_r^{(L)} \mathcal{W}_k$\;
    $\mathcal{V}_{r} \leftarrow \mathcal{E}_r^{(L)} \mathcal{W}_v$\;
    $\widetilde{\mathcal{E}}_r \leftarrow \mathcal{E}_r^{(L)} + \sum_{r^\prime}{Softmax\left(\frac{\mathcal{Q}_{r}\mathcal{K}_{r^\prime}^T}{\sqrt{d}}\right)\mathcal{V}_{r^\prime}}$\;
}
$\mathcal{E}^* \leftarrow Sum(\widetilde{\mathcal{E}}_1,...,\widetilde{\mathcal{E}}_R)$\;
Select ${\mathbf{e}_u^*}, {\mathbf{e}_i^*}$ from $\mathcal{E}^*$\;
$\widetilde{y}_{u,i} \leftarrow {\mathbf{e}_u^*}{\mathbf{e}_i^*}^T$\;
\end{algorithm}

\subsection{Model Complexity Analysis}
This part delves into the complexity analysis of SWGCN. As shown in Algorithm~\ref{algorithm}, the computational cost of SWGCN primarily arises from three components. First, the computation of interaction weights incurs $O(N_r \times d \times R)$, where $R$ typically assumes a small value ($R \ll N_r$), leading to an overall cost of $O(N_r \times d)$. Second, the encoding of graph networks entails a cost for multi-behavior graph encoding of $O(R \times L \times N_r \times d)$. Like the previous module, $R$ and $L$ typically assume the small values, and the final cost simplifies to $O(N_r \times d)$. Lastly, the cost of self-attention is $O((N_u + N_i) \times R^2 \times d)$. As the number of behaviors $R$ is significantly smaller than the node count ($R \ll N_u + N_i$), the aggregate cost of SWGCN is $O((N_u + N_i) \times d)$. Consequently, the computational cost of SWGCN scales as $O((N_r + N_u + N_i) \times d)$, which is consistent with the efficiency profile of typical GCN-based recommendation systems.

\section{Experiments}

We evaluate the proposed SWGCN model through rigorous testing on three distinct datasets, aiming to highlight its effectiveness and significant improvements over existing methods.

\subsection{Experimental Setup}

\subsubsection{Datasets} 

For the purpose of assessing our methodology, empirical trials are executed across a trio of unique data repositories gathered from actual scenarios, the specific characteristics of which are outlined in the following subsection.

\textbf{Taobao} \footnote{\href{https://tianchi.aliyun.com/dataset/dataDetail?dataId=649}{https://tianchi.aliyun.com/dataset/dataDetail?dataId=649}}: Taobao operates as a leading online marketplace in China, capturing four distinct categories of shopping behaviors: view, favorite, add to cart, and purchase.

\textbf{IJCAI} \footnote{\href{https://tianchi.aliyun.com/dataset/dataDetail?dataId=42}{https://tianchi.aliyun.com/dataset/dataDetail?dataId=42}}: This dataset, provided by the IJCAI competition, includes four types of interactions similar to the Taobao dataset and is characterized by a non-uniform distribution of multi-behavior interactions.

\textbf{Beibei} \footnote{\href{https://www.beibei.com}{https://www.beibei.com}}: Beibei is a prominent online retail platform in China that specializes in baby products, and its dataset comprises three distinct categories of shopping behaviors, i.e., view, add to cart, and purchase.

\begin{table*}[!ht]
    \centering
    \caption{Statistical information of three datasets.}
    \begin{tabular}{lrrr}
    \hline
        Dataset & Taobao & IJCAI & Beibei \\ \hline
        \#User & 111416 & 94849 & 21716 \\ 
        \#Item & 77321 & 51880 & 7977 \\ \hline
        \#view & 3937703 (81.87\%) & 6377858 (79.59\%) & 2412586 (72.75\%) \\ 
        \#Favorite & 176588 (3.67\%) & 713615 (8.91\%) & / \\ 
        \#Add to Cart & 448375 (9.32\%) & 9504 (0.12\%) & 642622 (19.38\%) \\ 
        \#Purchase & 247007 (5.14\%) & 912464 (11.39\%) & 261144 (7.87\%) \\ \hline
        \#Total Interactions & 4809673 & 8013441 & 3316352 \\ \hline
    \end{tabular}
    \label{tab: dataset info}
\end{table*}

\begin{table*}[!ht]
    \centering
    \caption{Comparison of the overall performance of all baseline methods on Taobao. \textbf{Bolded} numbers denote superior performance metrics, whereas \underline{underlined} numbers signify sub-optimal performance. SWGCN-T denotes that interaction weights for the target behavior are concurrently trained via the Synergy Alignment Task. H@K signifies HR@K and N@K denotes NDCG@K.}
    \begin{tabular}{lcccccccccc}
    \hline
         Models & H@10 & N@10 & H@20 & N@20 & H@50 & N@50 & H@100 & N@100 & H@200 & N@200 \\ \hline
         BPR-MF~\cite{bprmf} & 0.0417 & 0.0224 & 0.0628 & 0.0277 & 0.1189 & 0.0387 & 0.2048 & 0.0526 & 0.3393 & 0.0713 \\ 
         NCF-GMF~\cite{ncf} & 0.0414 & 0.0212 & 0.0647 & 0.0271 & 0.1130 & 0.0366 & 0.1712 & 0.0460 & 0.2648 & 0.0590 \\
         NCF-MLP~\cite{ncf} & 0.0365 & 0.0201 & 0.0554 & 0.0248 & 0.1120 & 0.0358 & 0.2104 & 0.0516 & 0.3431 & 0.0701 \\ 
         NCF-NeuMF~\cite{ncf} & 0.0373 & 0.0204 & 0.0559 & 0.0250 & 0.1114 & 0.0360 & 0.2080 & 0.0516 & 0.3431 & 0.0704 \\
         NGCF-OB~\cite{ngcf} & 0.0352 & 0.0203 & 0.0587 & 0.0262 & 0.1165 & 0.0375 & 0.1994 & 0.0509 & 0.3261 & 0.0685 \\ 
         LightGCN~\cite{lightgcn} & 0.0370 & 0.0213 & 0.0571 & 0.0263 & 0.1059 & 0.0359 & 0.1921 & 0.0498 & 0.3304 & 0.0691 \\ \hline
         NGCF-MB~\cite{ngcf} & 0.0059 & 0.0029 & 0.0104 & 0.0040 & 0.0203 & 0.0060 & 0.0332 & 0.0080 & 0.0529 & 0.0108 \\ 
         NMTR-GMF~\cite{nmtr} & 0.0396 & 0.0207 & 0.0634 & 0.0267 & 0.1166 & 0.0371 & 0.1977 & 0.0502 & 0.3283 & 0.0683 \\ 
         MBGCN~\cite{mbgcn} & 0.0471 & 0.0257 & 0.0719 & 0.0320 & 0.1351 & 0.0444 & 0.2117 & 0.0567 & 0.3295 & 0.0731 \\ 
         GHCF~\cite{ghcf} & 0.0419 & 0.0197 & 0.0682 & 0.0263 & 0.1251 & 0.0374 & 0.2001 & 0.0495 & 0.3365 & 0.0684 \\ \hline
         SWGCN (Ours) & \textbf{0.1328} & \textbf{0.0809 }& \textbf{0.1832} & \textbf{0.0935} & \textbf{0.2766} & \textbf{0.1120} & \underline{0.3713} & \underline{0.1273} & \underline{0.4951} & \underline{0.1446} \\ 
         SWGCN-T (Ours) & \underline{0.1313} & \underline{0.0804} & \underline{0.1817} & \underline{0.0931} & \underline{0.2764} & \underline{0.1118} & \textbf{0.3730} & \textbf{0.1274} & \textbf{0.4967} & \textbf{0.1447} \\ \hline
    \end{tabular}
    \label{tab: performance on taobao}
\end{table*}

We designate purchasing as the principal objective in these datasets; conversely, the rest of the user activities are categorized as secondary information. To ensure the integrity of the datasets, we adhere to the configurations established by related methodologies~\cite{ghcf, chen2020efficienta, chen2020efficientb}. Initially, redundant user-item interactions are consolidated by retaining the foremost interaction instances. For Taobao and IJCAI, we filter out entities with fewer than five buying records, whereas Beibei is kept raw. A summary of these final datasets appears in ~\ref{tab: dataset info}. We adopt a temporal split where the last transaction is utilized for the test phase, the second-to-last record is reserved for validation purposes, and the prior history constitutes the training set.

\subsubsection{Evaluation Metrics}

We employ a pair of widely-recognized indices to gauge the efficacy of Top-$K$ ranking outcomes, Normalized Discounted Cumulative Gain (NDCG) and Hit Ratio (HR)~\cite{xia2021multi, xia2021graph, ghcf}. The specific formulations of these measures are presented below:

\begin{itemize}[leftmargin=*]
    \item \textbf{HR@K} measures the proportion of Top-K recommendation lists that successfully include recommended items, thereby indicating the accuracy of successful recommendations.
    \item \textbf{NDCG@K} accounts for position sensitivity by assigning higher weights to relevant items appearing near the top of the recommendation queue, thereby emphasizing the quality of the ranking order.
\end{itemize}

Significantly, the assessment involves ordering all items per user, a practice more aligned with the demands of actual recommendation environments.

\begin{table*}[!ht]
    \centering
    \caption{Comparison of the overall performance of all baseline methods on IJCAI. \textbf{Bolded} numbers denote superior performance metrics, whereas \underline{underlined} numbers signify sub-optimal performance. SWGCN-T denotes that interaction weights for the target behavior are concurrently trained via the Synergy Alignment Task. H@K signifies HR@K and N@K denotes NDCG@K.}
    \begin{tabular}{lcccccccccc}
    \hline
        Models & H@10 & N@10 & H@20 & N@20 & H@50 & N@50 & H@100 & N@100 & H@200 & N@200 \\ \hline
        BPR-MF~\cite{bprmf} & 0.0097 & 0.0051 & 0.0152 & 0.0065 & 0.0278 & 0.0089 & 0.0437 & 0.0115 & 0.0694 & 0.0151 \\ 
        NCF-GMF~\cite{ncf} & 0.0103 & 0.0054 & 0.0162 & 0.0069 & 0.0297 & 0.0095 & 0.0469 & 0.0123 & 0.0738 & 0.0161 \\ 
        NCF-MLP~\cite{ncf} & 0.0085 & 0.0043 & 0.0133 & 0.0055 & 0.0260 & 0.0080 & 0.0419 & 0.0106 & 0.0674 & 0.0141 \\ 
        NCF-NeuMF~\cite{ncf} & 0.0087 & 0.0044 & 0.0143 & 0.0058 & 0.0260 & 0.0081 & 0.0422 & 0.0107 & 0.0679 & 0.0143 \\ 
        NGCF-OB~\cite{ngcf} & 0.0103 & 0.0050 & 0.0174 & 0.0068 & 0.0336 & 0.0100 & 0.0539 & 0.0133 & 0.0852 & 0.0177 \\ 
        LightGCN~\cite{lightgcn} & 0.0094 & 0.0048 & 0.0148 & 0.0062 & 0.0263 & 0.0084 & 0.0437 & 0.0113 & 0.0699 & 0.0149 \\ \hline
        NGCF-MB~\cite{ngcf} & 0.0120 & 0.0060 & 0.0203 & 0.0081 & \underline{0.0392} & 0.0118 & \textbf{0.0636} & \textbf{0.0158} & \textbf{0.1006} & \textbf{0.0209} \\ 
        NMTR-GMF~\cite{nmtr} & 0.0111 & 0.0058 & 0.0171 & 0.0073 & 0.0306 & 0.0100 & 0.0478 & 0.0128 & 0.0742 & 0.0165 \\ 
        MBGCN~\cite{mbgcn} & 0.0109 & 0.0056 & 0.0165 & 0.0071 & 0.0295 & 0.0096 & 0.0463 & 0.0123 & 0.0721 & 0.0159 \\ 
        GHCF~\cite{ghcf} & 0.0064 & 0.0037 & 0.0123 & 0.0052 & 0.0225 & 0.0072 & 0.0371 & 0.0095 & 0.0614 & 0.0129 \\ \hline
        SWGCN (Ours) & \underline{0.0126} & \underline{0.0064} & \textbf{0.0208} & \textbf{0.0084} & \textbf{0.0393} & \textbf{0.0120} & \underline{0.0622} & \underline{0.0157} & \underline{0.0958} & \underline{0.0204} \\ 
        SWGCN-T (Ours) & \textbf{0.0127} & \textbf{0.0064} & \underline{0.0207} & \underline{0.0084} & 0.0387 & \underline{0.0119} & 0.0616 & 0.0156 & 0.0949 & 0.0203 \\ \hline
    \end{tabular}
    \label{tab: performance on ijcai}
\end{table*}

\begin{table*}[!ht]
    \centering
    \caption{Comparison of the overall performance of all baseline methods on Beibei. \textbf{Bolded} numbers denote superior performance metrics, whereas \underline{underlined} numbers signify sub-optimal performance. SWGCN-T denotes that interaction weights for the target behavior are concurrently trained via the Synergy Alignment Task. H@K signifies HR@K and N@K denotes NDCG@K.}
    \begin{tabular}{lcccccccccc}
    \hline
        Models & H@10 & N@10 & H@20 & N@20 & H@50 & N@50 & H@100 & N@100 & H@200 & N@200 \\ \hline
        BPR-MF~\cite{bprmf} & 0.0343 & \textbf{0.0188} & 0.0476 & \textbf{0.0221} & 0.0854 & 0.0295 & 0.1274 & 0.0362 & 0.2491 & 0.0531 \\ 
        NCF-GMF~\cite{ncf} & 0.0345 & 0.0161 & 0.0481 & 0.0196 & 0.0834 & 0.0266 & 0.1234 & 0.0330 & 0.2472 & 0.0500 \\ 
        NCF-MLP~\cite{ncf} & 0.0261 & 0.0107 & 0.0353 & 0.0130 & 0.0635 & 0.0185 & 0.1100 & 0.0261 & 0.2396 & 0.0441 \\ 
        NCF-NeuMF~\cite{ncf} & 0.0244 & 0.0095 & 0.0331 & 0.0117 & 0.0636 & 0.0175 & 0.1131 & 0.0255 & 0.2449 & 0.0438 \\ 
        NGCF-OB~\cite{ngcf} & 0.0299 & 0.0155 & 0.0417 & 0.0185 & 0.0797 & 0.0259 & 0.1154 & 0.0317 & 0.2242 & 0.0466 \\ 
        LightGCN~\cite{lightgcn} & 0.0323 & \underline{0.0186} & 0.0436 & 0.0214 & 0.0768 & 0.0279 & 0.1135 & 0.0338 & 0.2394 & 0.0512 \\  \hline
        NGCF-MB~\cite{ngcf} & \textbf{0.0366} & 0.0170 & \underline{0.0559} & \underline{0.0219} & \textbf{0.1140} & \textbf{0.0333} & 0.1763 & \underline{0.0434} & \underline{0.2739} & \underline{0.0569} \\ 
        NMTR-GMF~\cite{nmtr} & 0.0257 & 0.0134 & 0.0418 & 0.0175 & 0.0768 & 0.0244 & 0.1219 & 0.0316 & 0.2361 & 0.0474 \\ 
        MBGCN~\cite{mbgcn} & 0.0322 & 0.0174 & 0.0419 & 0.0198 & 0.0793 & 0.0272 & 0.1258 & 0.0346 & 0.2439 & 0.0509 \\ 
        GHCF~\cite{ghcf} & \underline{0.0357} & 0.0124 & 0.0530 & 0.0167 & 0.0867 & 0.0233 & 0.1408 & 0.0320 & 0.2321 & 0.0448 \\ \hline
        SWGCN (Ours) & 0.0315 & 0.0150 & \textbf{0.0564} & 0.0213 & \underline{0.1119} & \underline{0.0322} & \underline{0.1774} & 0.0428 & 0.2730 & 0.0561 \\ 
        SWGCN-T (Ours) & 0.0317 & 0.0152 & 0.0556 & 0.0212 & 0.1108 & 0.0320 & \textbf{0.1824} & \textbf{0.0436} & \textbf{0.2822} & \textbf{0.0575} \\ \hline
    \end{tabular}
    \label{tab: performance on beibei}
\end{table*}

\subsubsection{Baselines}
To illustrate the efficacy of SWGCN, a comprehensive comparative analysis is conducted with a range of baselines. We categorize the comparison models into two main classes: single-behavior approaches and multi-behavior frameworks. The former relies exclusively on the primary interaction data, whereas the latter incorporates diverse behavioral signals.

\textbf{Single-behavior Methods}:

\begin{itemize}[leftmargin=*]
    \item \textbf{BPR-MF}~\cite{bprmf}: BPR is a widely used criterion to optimize pairwise training loss. This study compares a BPR-informed matrix factorization model, referred to as \textbf{BPR-MF}.
    \item \textbf{NCF}~\cite{ncf}: NCF operates as a neural architecture for recommendation. We evaluate three distinct configurations of this framework: the version utilizing Generalized Matrix Factorization, the version based on Multi-Layer Perceptrons, and the NeuMF architecture, which represents an integration of both. These are labeled as \textbf{NCF-GMF}, \textbf{NCF-MLP}, and \textbf{NCF-NeuMF}, respectively.
    \item \textbf{NGCF-OB}~\cite{ngcf}: NGCF is an advanced graph neural network model that enables high-order message propagation on user-item bipartite graphs. In this specific setting, we build the model by leveraging only the target behavior to generate the interaction network, which is named \textbf{NGCF-OB}.
    \item \textbf{LightGCN}~\cite{lightgcn}: a streamlined iteration of NGCF, utilizes a simplified architecture for more effective and efficient collaborative filtering.
\end{itemize}

\textbf{Multi-behavior Methods}:

\begin{itemize}[leftmargin=*]
    \item \textbf{NGCF-MB}~\cite{ngcf}: In a departure from \textbf{NGCF-OB}, this model incorporates all categories of user activities to form the user-item graph. However, these varied interactions are consolidated into a singular type, creating a graph with homogeneous edges.
    \item \textbf{NMTR}~\cite{nmtr}: NMTR is an advanced multi-behavior model that employs multi-task learning to leverage diverse behavior data. It integrates NCF for each behavior within a multi-task learning framework, utilizing GMF here, denoted as \textbf{NMTR-GMF}.
    \item \textbf{MBGCN}~\cite{mbgcn}: This method represents a recommendation technique that constructs uniform heterogeneous graphs to model multi-behavior interactions and employs learnable weights to capture the influence of distinct behaviors.
     \item \textbf{GHCF}~\cite{ghcf}: GHCF is a multi-behavior recommendation approach that omits negative sampling, focusing on high-hop heterogeneous user-item interactions.
\end{itemize}

\subsubsection{Parameter Settings}

The SWGCN framework is built upon PyTorch. For optimization, the Adam optimizer~\cite{kingma2014adam} is employed for every model, utilizing a uniform training batch of 2048. Furthermore, a consistent embedding dimensionality of 32 is maintained across all experiments. For sampling-based approaches, the negative sampling count is established at 4, an empirically determined value that yields favorable performance. Parameters are initialized using Xavier initialization~\cite{glorot2010understanding}. During the training phase, all comparative models adhere to the optimal configurations specified in their original publications. An exhaustive grid search is performed for SWGCN, with the learning rate examined across the interval of $\{1 \times 10^{-5}, 5 \times 10^{-5}, 1 \times 10^{-4}, 5 \times 10^{-4}, 1 \times 10^{-3}\}$, with the optimal learning rate identified as \(1 \times 10^{-3}\). Additionally, the \(L_2\) regularization coefficient was examined within the interval $\{1 \times 10^{-5}, 1 \times 10^{-4}, 1 \times 10^{-3}, 1 \times 10^{-2}\}$, resulting in the optimal \(L_2\) regularization coefficient being established at \(1 \times 10^{-5}\) for both \(\gamma_1\) and \(\gamma_2\). The encoding graph network of SWGCN is set to $L=3$, based on empirical findings~\cite{ngcf, lightgcn}. Moreover, to prevent overfitting, an early stopping mechanism is utilized. The training procedure ceases when the HR@10 score on the validation data fails to improve over 50 consecutive epochs.

\subsection{Overall Performance Comparison}

In this section, the overall effectiveness of SWGCN is evaluated against several baseline methods. Each model is initialized with random values and executed five times, with the average of the outcomes serving as the definitive measure of its performance. The findings on the three real-world datasets are detailed in~\ref{tab: performance on taobao},~\ref{tab: performance on ijcai}, and~\ref{tab: performance on beibei}. From the experimental findings, we draw the subsequent conclusions.

On the Taobao dataset, SWGCN's performance is superior across all metrics, substantially surpassing the established state-of-the-art (SOTA) benchmarks. To be specific, when compared against the strongest baseline, SWGCN provides an average enhancement of 112.49\% for HR metrics and a 156.36\% boost for NDCG metrics. This empirical evidence robustly supports the effectiveness of SWGCN, which integrates the Target Preference Weigher and Synergy Alignment Task, in enhancing recommendation outcomes.

Regarding the IJCAI and Beibei datasets, SWGCN achieves the highest performance on partial metrics. In particular, SWGCN enhances fine-grained recommendations on the IJCAI dataset, yielding an average increase of 2.77\% in HR metrics and 3.77\% in NDCG metrics at K values of 10, 20, and 50. Similarly, on the Beibei dataset, SWGCN improves the performance of coarse-grained recommendations, achieving an average enhancement of 3.02\% in HR and 0.75\% in NDCG at K values of 100 and 200. These results indicate that SWGCN outperforms existing SOTA methods on both the IJCAI and Beibei datasets, thereby substantiating its capacity to generalize across various recommendation scenarios to a certain extent.

Additionally, we introduce a variant framework, SWGCN-T, which utilizes the Synergy Alignment Task to support the training of the Target Preference Weigher with the target behavior. This variant achieves results comparable to the base SWGCN on both Taobao and IJCAI. However, on the Beibei dataset, it shows an improvement in coarse-grained recommendation performance, particularly for HR and NDCG at K=100 and 200. Overall, integrating the Synergy Alignment Task to assist in training interaction weights for the target behavior has a negligible effect on model performance. This observation substantiates the hypothesis that the proposed Synergy Alignment Task primarily leverages synergistic signals connecting auxiliary actions to the target action, while deriving limited benefit from co-occurrence information of the target action itself.

Moreover, the analysis of overall performance reveals several intriguing insights. First, on the IJCAI and Beibei datasets, certain complex multi-behavior models do not outperform simple multi-behavior aggregation methods, such as NGCF-MB~\cite{ngcf}. A plausible explanation for this phenomenon is the diminished robustness of contemporary multi-behavior recommendation methodologies, which require meticulous hyperparameter tuning when datasets are altered. Second, the performance improvement of SWGCN on IJCAI and Beibei is less pronounced compared to Taobao. The underlying factors contributing to this discrepancy warrant further investigation.

\begin{figure*}[pos=!t]
    \centering
    \subfigure[HR comparison of ablation experiment results on Taobao.]{
        \includegraphics[width=0.45\linewidth]{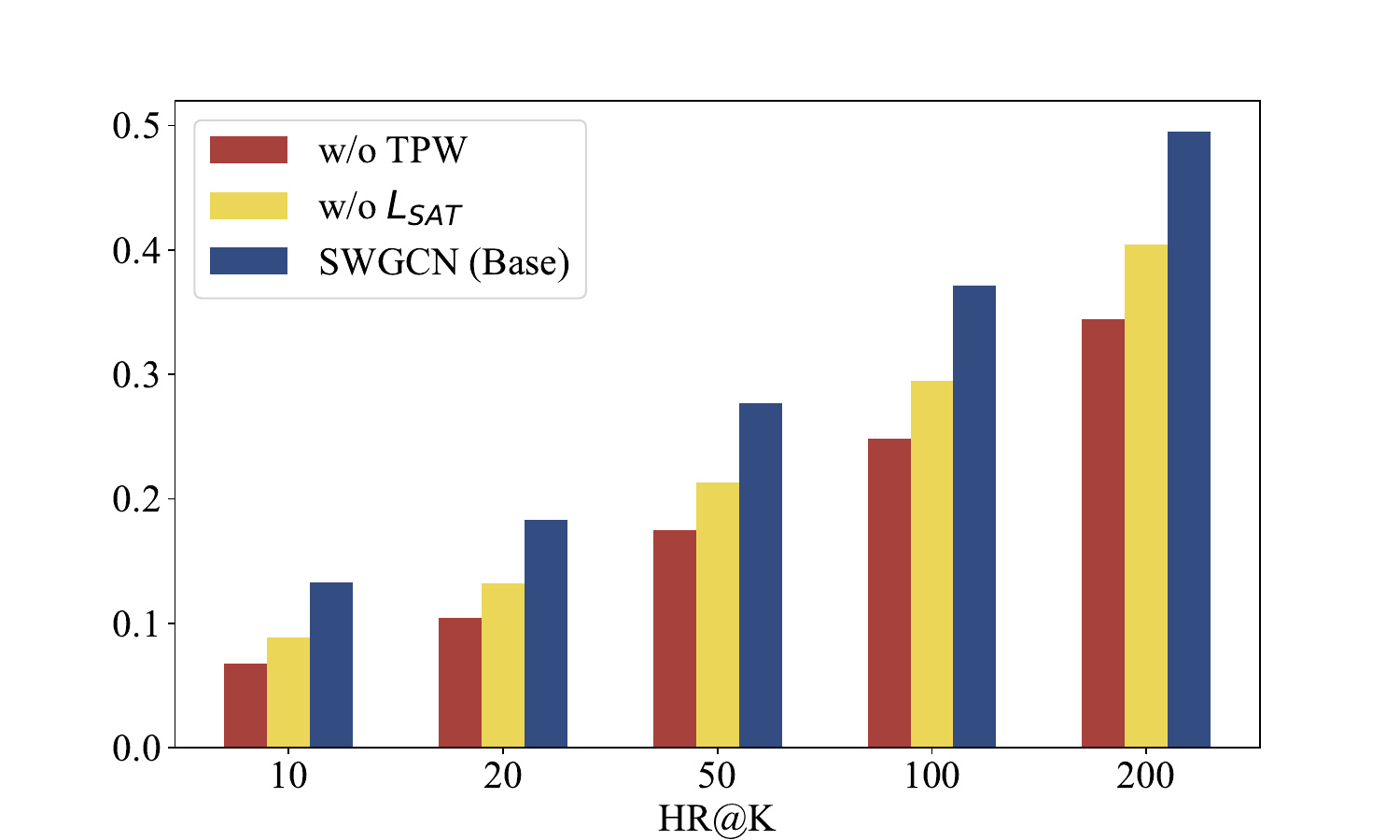}
    }
    \subfigure[NDCG comparison of ablation experiment results on Taobao.]{
	\includegraphics[width=0.45\linewidth]{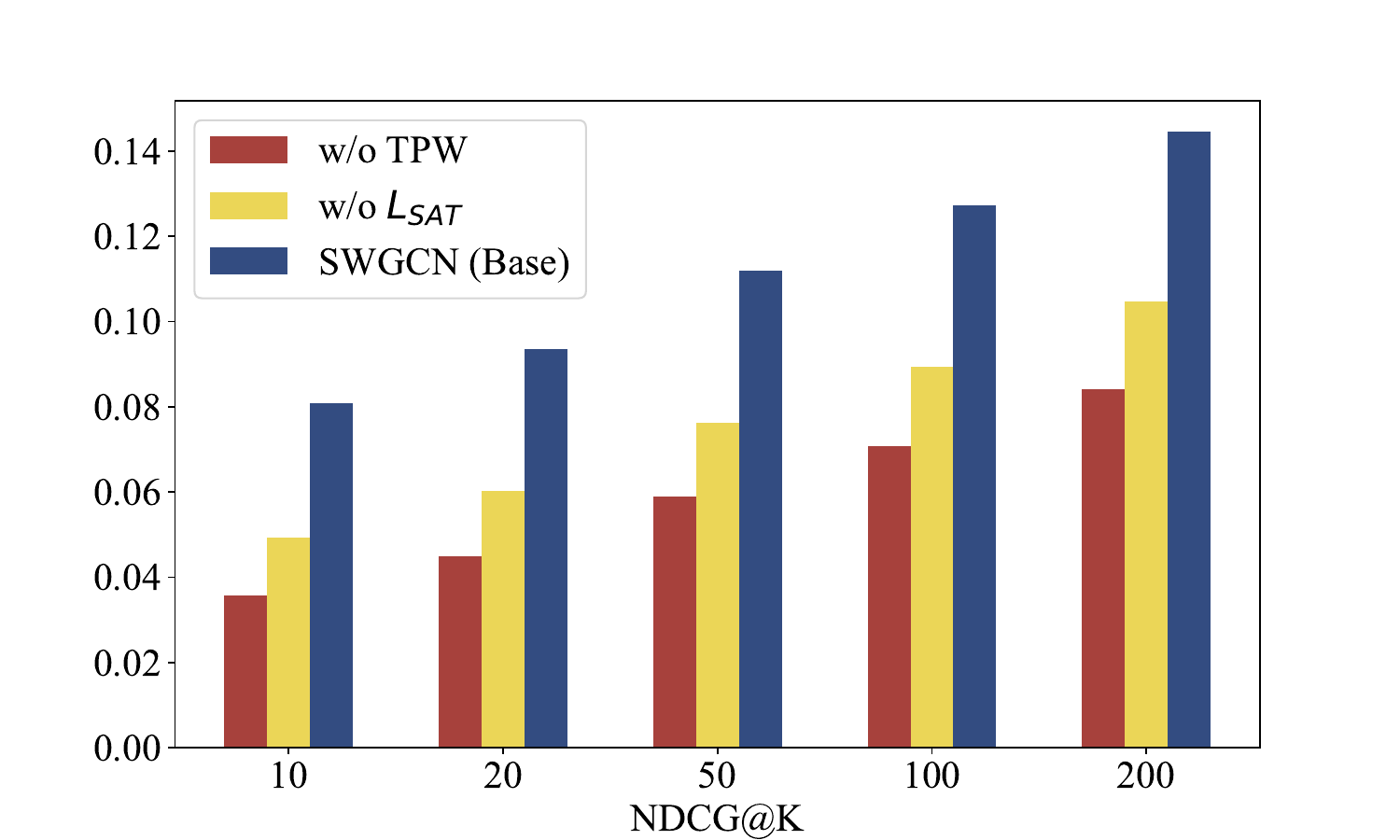}
    }
    \quad    
    \subfigure[HR comparison of ablation experiment results on Beibei.]{
    	\includegraphics[width=0.45\linewidth]{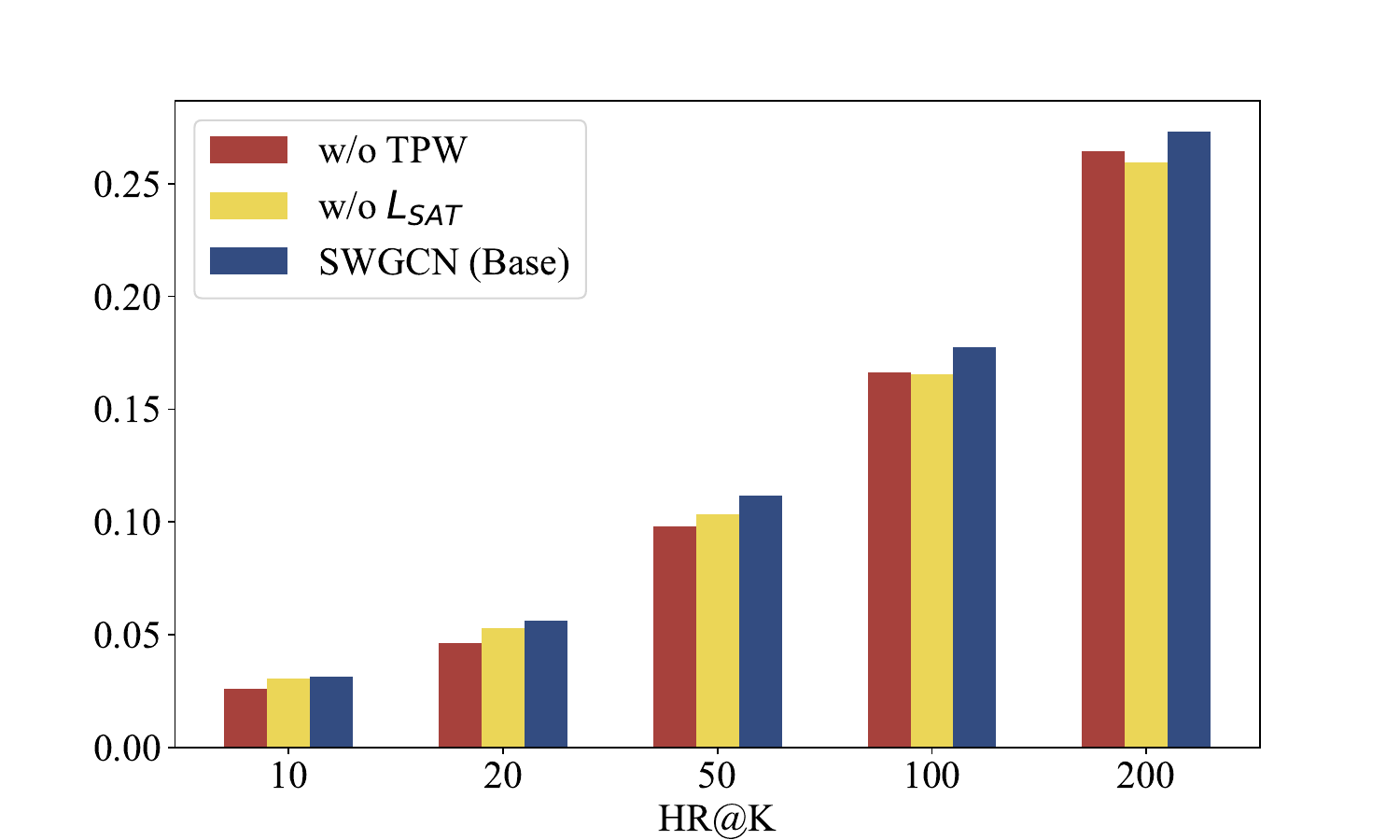}
    }
    \subfigure[NDCG comparison of ablation experiment results on Beibei.]{
	\includegraphics[width=0.45\linewidth]{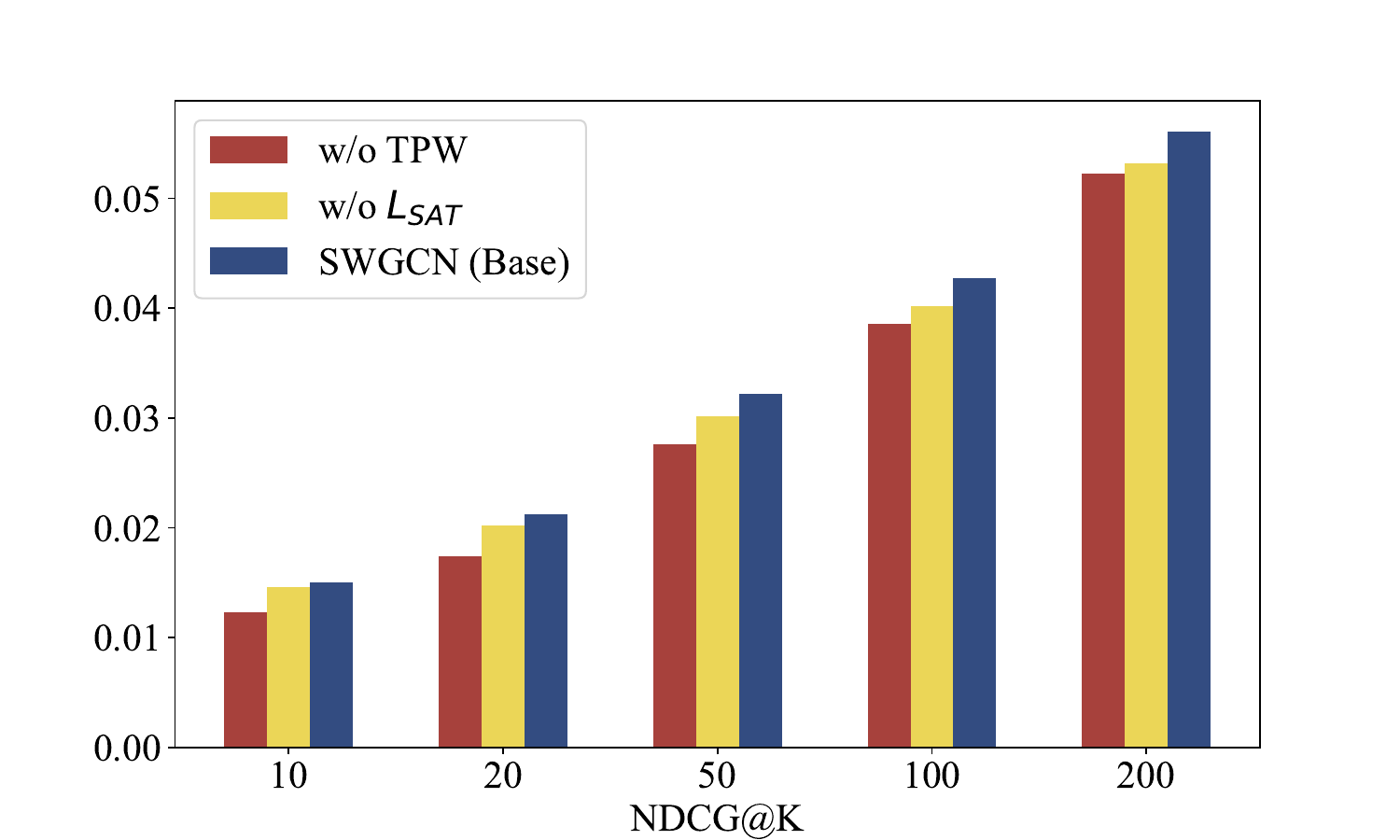}
    }
    \caption{Visualization of ablation experiment results.}
    \label{fig: ablation study}
    \vspace{-0.5cm}
\end{figure*}

\subsection{Ablation Study} 

To comprehensively analyze the contribution of each sub-module within SWGCN, two variant models are developed for comparative analysis as follows:
\begin{itemize}[leftmargin=*]
\item ``w/o $L_{SAT}$'': This variant omits the Synergy Alignment Task during the training process, relying solely on the BPR Loss for optimization. 
\item ``w/o TPW'': This variant eliminates the Target Preference Weigher, with training solely on the BPR Loss. 
\end{itemize}

Each experimental result is repeated five times, and the mean value is recorded as the ultimate performance. The findings from the ablation analysis are depicted in~\ref{fig: ablation study}. We draw some conclusions from the experimental results.

Preliminary results indicate that the Synergy Alignment Task significantly improves model performance. For the Taobao dataset, a performance degradation is observed when this task is omitted, with an average reduction of 24.70\% in the HR metric and 32.79\% in the NDCG metric. Likewise, on the Beibei dataset, there are average reductions of 5.69\% and 4.92\% in the HR and NDCG metrics, respectively. The empirical findings substantiate the efficacy of the proposed Synergy Alignment Task, demonstrating that leveraging the collaborative interplay between supplementary and primary user actions can significantly elevate the effectiveness of recommendation engines.

Furthermore, Target Preference Weigher is essential for improving model performance. In stark contrast to the baseline SWGCN model, the ``w/o TPW'' variant exhibits a pronounced deterioration in recommendation efficacy. On the Taobao dataset, there is an average performance decrease of 38.53\% in HR and 48.23\% in NDCG. Similarly, for the Beibei dataset, the variant shows an average reduction in effectiveness of 11.44\% for HR and 13.35\% for NDCG.

\begin{figure*}[pos=!t]
    \centering
    \subfigure[Results of different $\lambda_s$ on Taobao.]{
        \includegraphics[width=0.45\linewidth]{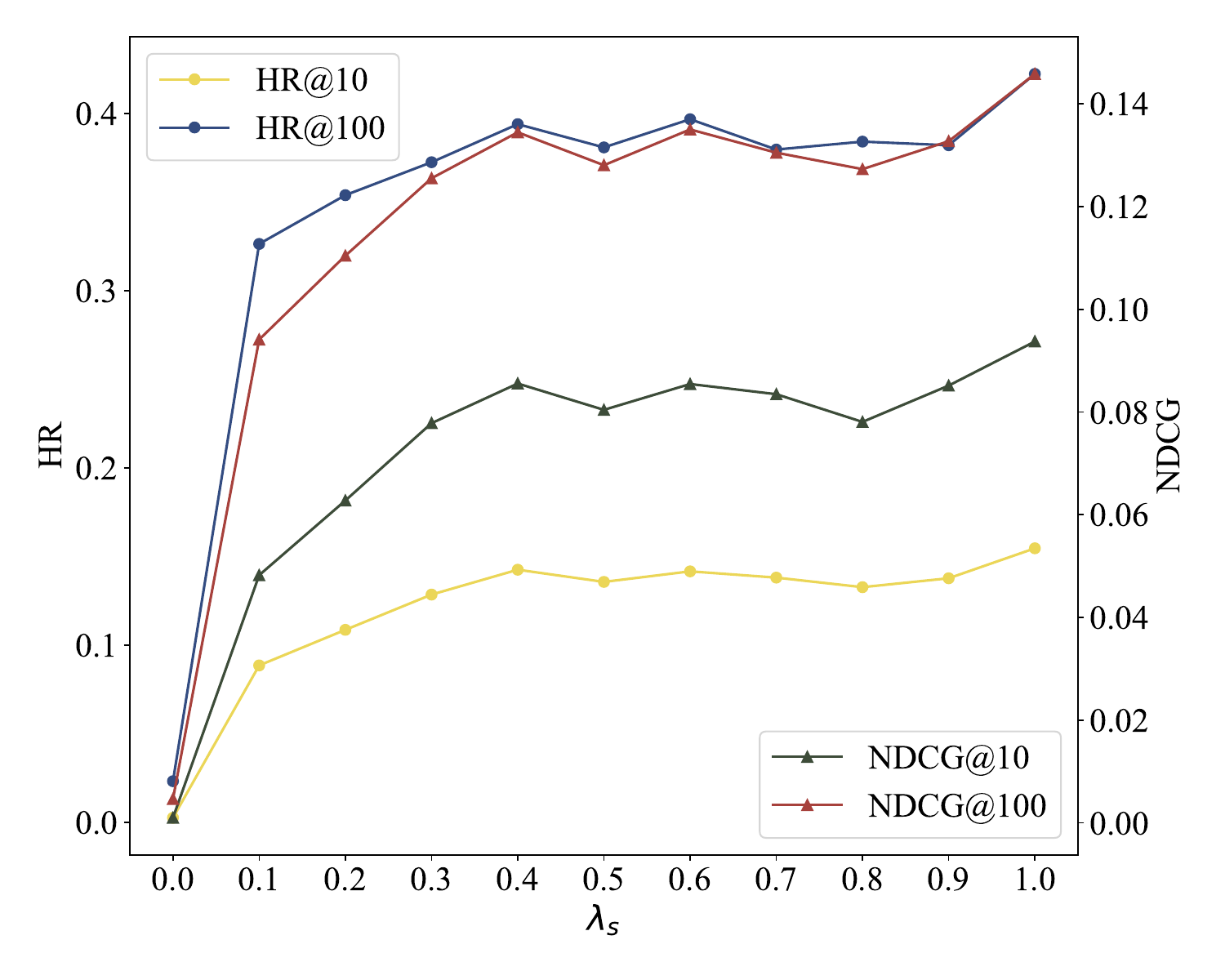}
    }
    \subfigure[Results of different $\lambda_s$ on Beibei.]{
	\includegraphics[width=0.45\linewidth]{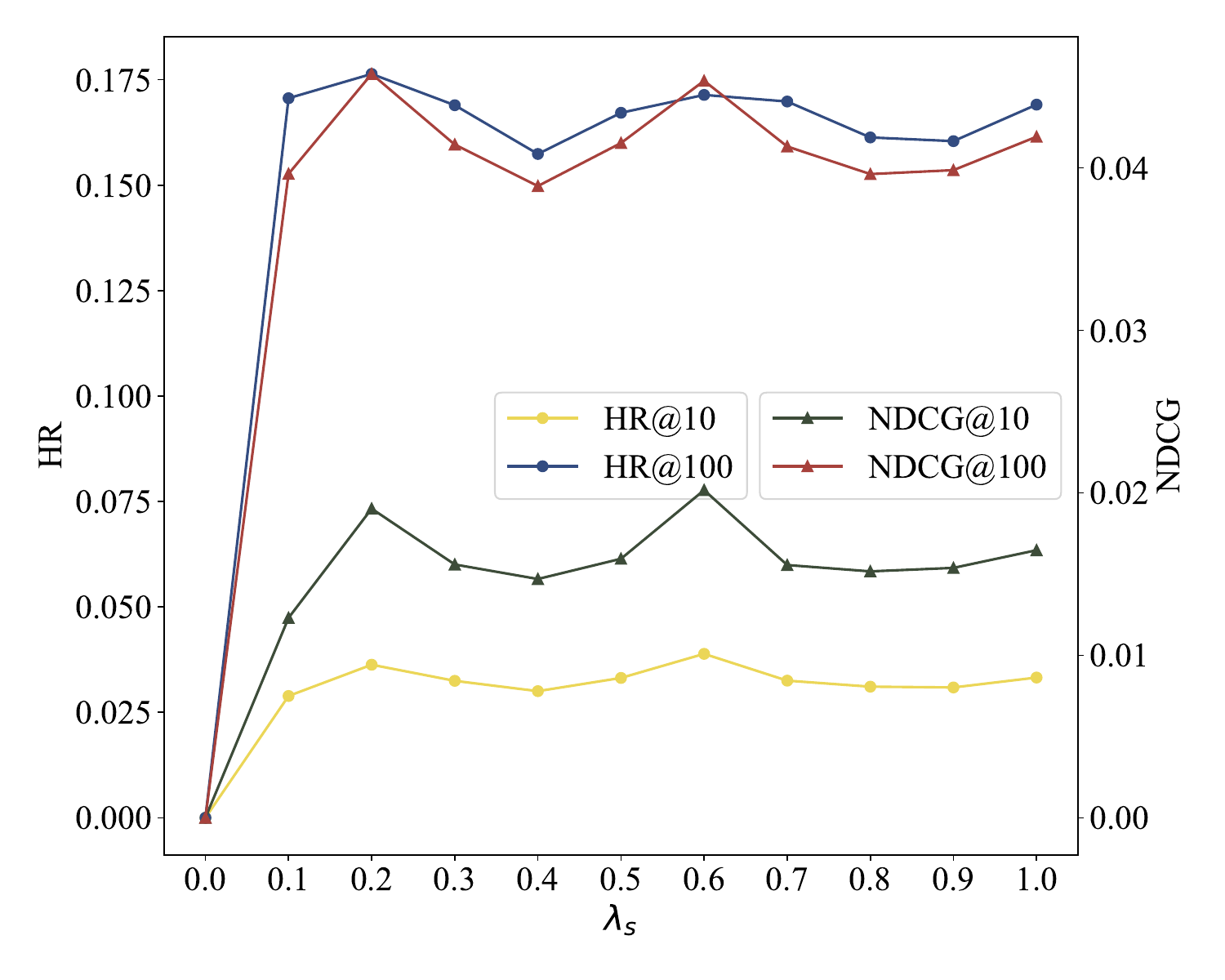}
    }
    \caption{Visualization of self-loop weights affecting results.}
    \label{fig: self loop performance}
    \vspace{-0.5cm}
\end{figure*}

\begin{figure*}[pos=!t]
    \subfigure[Distribution of interaction weights for different behaviors on Taobao.]{
    	\includegraphics[width=0.46\linewidth]{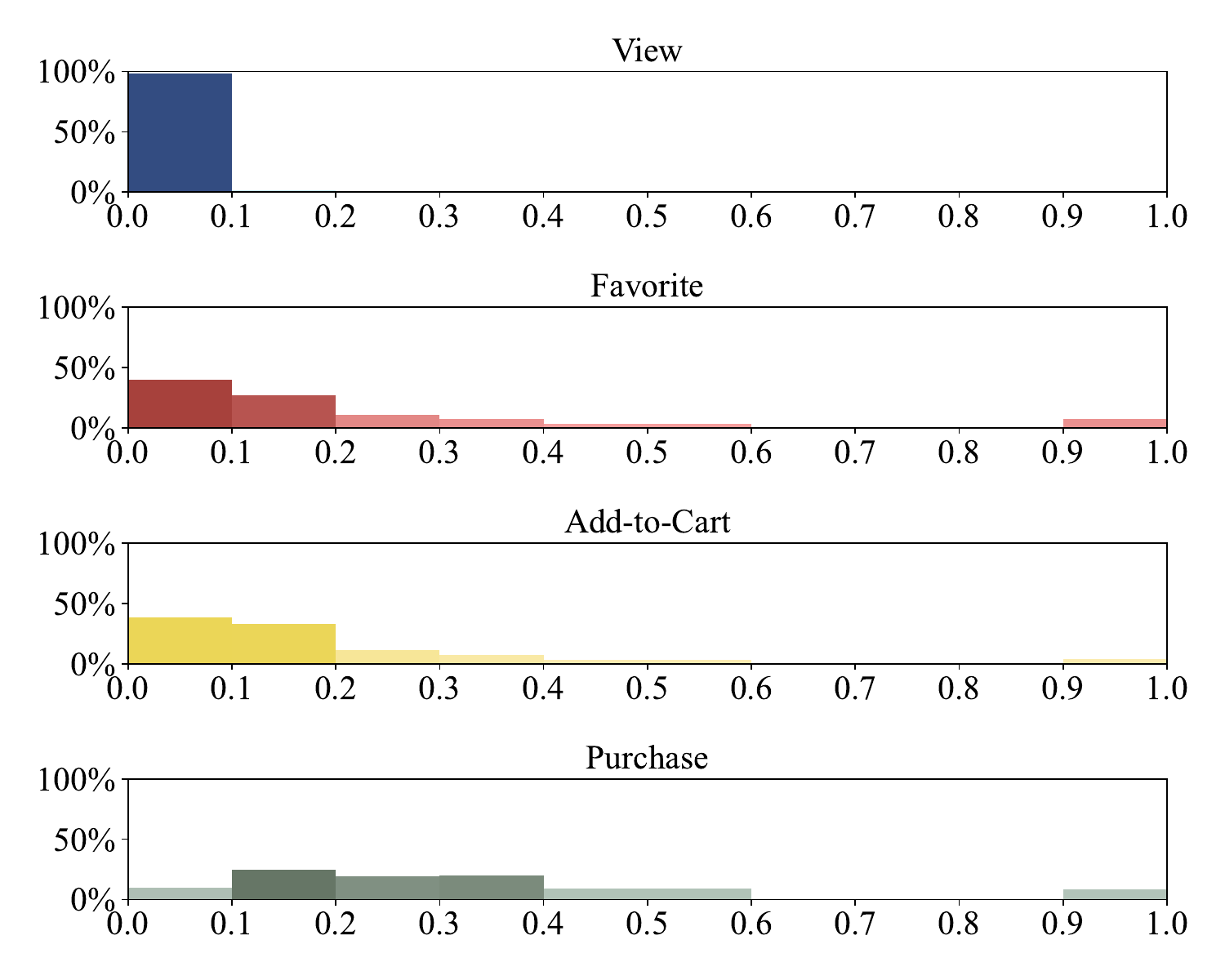}
    }
    \subfigure[Distribution of interaction weights for different behaviors on Beibei.]{
	\includegraphics[width=0.46\linewidth]{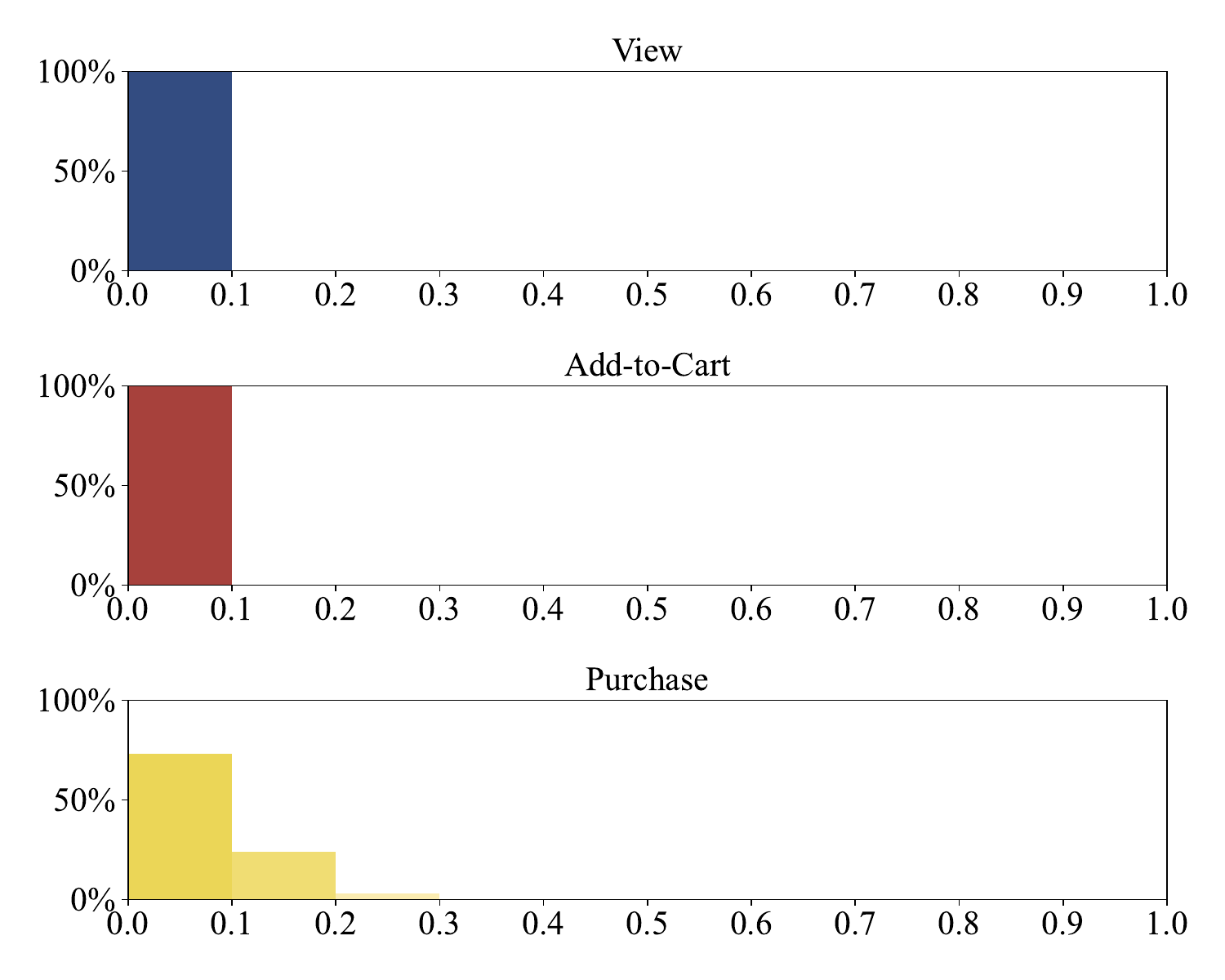}
    }
    \caption{Visualization of the distribution of interaction weights for different behaviors.}
    \label{fig: interaction weight}
    \vspace{-0.5cm}
\end{figure*}

\subsection{Hyperparameter Analysis}

This section aims to provide practical guidance on parameter selection for SWGCN by analyzing the model's sensitivity to its key hyperparameters, thus addressing how to set the parameters for better performance. The primary hyperparameters of SWGCN are categorized into three dimensions: the self-loop weight, the weight associated with the Synergy Alignment Task, and the message dropout rate. We analyze the results of these three aspects of the experiments in the following subsections. Note that the performance metrics reported are derived from a single experimental run.

\subsubsection{Impact of Self-loop Weight} 
\label{self_loop_weight_exp}
To scrutinize the impact of the self-loop weight on the efficacy of SWGCN, we assess the framework's results across a spectrum of values for this weight, $\lambda_s$, ranging from 0.0 to 1.0 in increments of 0.1, utilizing the Taobao and Beibei datasets. The findings from this assessment are depicted in~\ref{fig: self loop performance}. Additionally, to elucidate the underlying influence mechanisms of the self-loop weights, we perform a statistical analysis of the interaction weight distribution across various behaviors in the datasets, with the findings presented in~\ref{fig: interaction weight}.

The analysis presented in~\ref{fig: self loop performance} demonstrates that nullifying the self-loop weight ($\lambda_s = 0$) results in a significant deterioration of recommendation efficacy, nearing minimal levels across both the Taobao and Beibei datasets. This observation underscores the indispensable contribution of self-loops to the graph network's encoding mechanism. As the self-loop weight is adjusted, the recommendation performance on the Beibei dataset shows minimal variation, with optimal performance occurring at $\lambda_s = 0.2$. Conversely, for the Taobao dataset, the recommendation performance initially increases before stabilizing, achieving peak performance at $\lambda_s = 1.0$.

~\ref{fig: interaction weight} reveals noteworthy patterns. For the Beibei dataset, interaction weights associated with auxiliary behaviors predominantly fall within the range of 0.0 to 0.1, while those corresponding to the target behavior are primarily distributed between 0.0 and 0.2. Combined with the earlier analysis, the optimal self-loop weight, $\lambda_s = 0.2$, this finding implies that a strong correspondence between this specific outcome and the interaction weights relevant to the primary user activity within the Beibei data. This observation implies that, within the Beibei dataset, the information of the node itself is nearly as critical as that of its neighboring nodes during graph network encoding, indicating a lower prevalence of interfering interactions.

In contrast, the Taobao dataset exhibits a different pattern. Here, interaction weights for auxiliary behaviors are mainly distributed within the range of 0.0 to 0.2. By comparison, interaction weights for the target behavior are predominantly found between 0.1 and 0.4, with a minority extending to the range of 0.9 to 1.0. Nevertheless, consistent with previous findings, the optimal self-loop weight for the Taobao dataset is identified as $\lambda_s = 1.0$. This result suggests a significant disparity between the self-loop weight and the interaction weights on the Taobao dataset, indicating that self-information is more critical than information from neighboring nodes in the graph network encoding. Consequently, this implies a higher incidence of interfering interactions and more tremendous noise within the Taobao dataset.

\begin{figure*}[pos=!t]
    \centering
    \subfigure[Results of different $\lambda_a$ on Taobao.]{
        \includegraphics[width=0.4\linewidth]{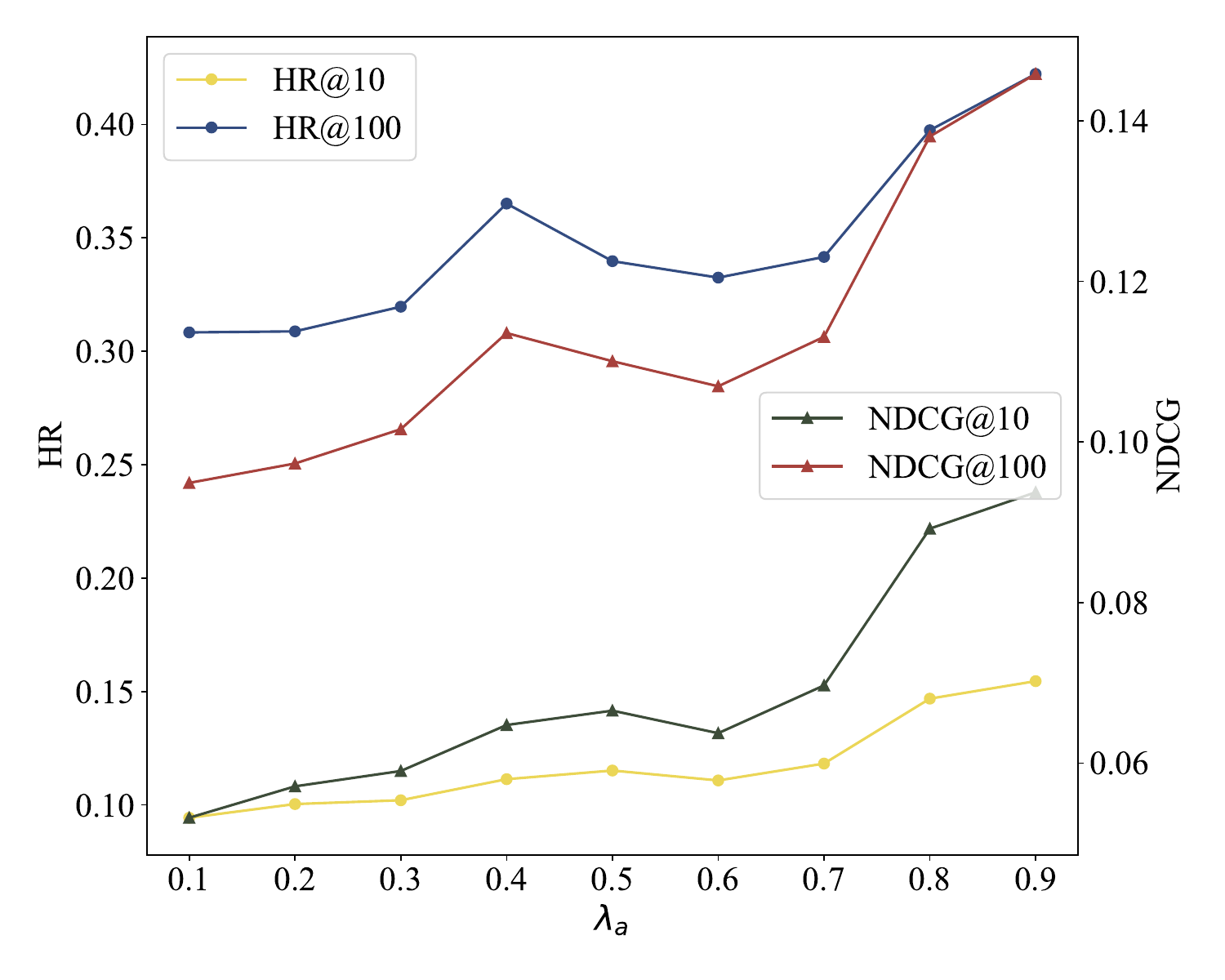}
    }
    \subfigure[Results of different $\lambda_a$ on Beibei.]{
	\includegraphics[width=0.4\linewidth]{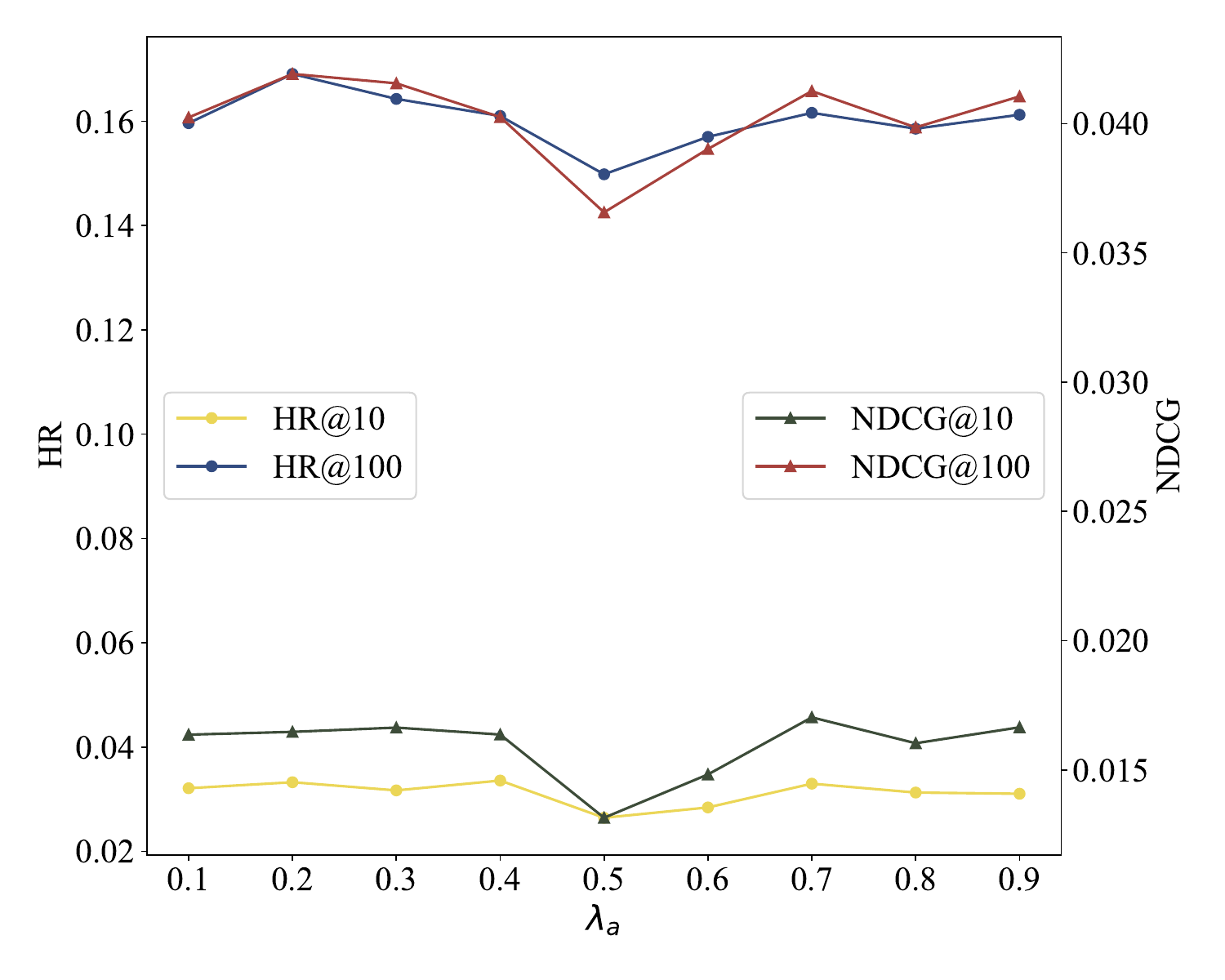}
    }
    \caption{Visualization of the impact results for the Synergy Alignment Task weights.}
    \label{fig: lambda performance}
    \vspace{-0.5cm}
\end{figure*}

\begin{figure}[pos=!t]
    \centering
	\includegraphics[width=0.4\linewidth]{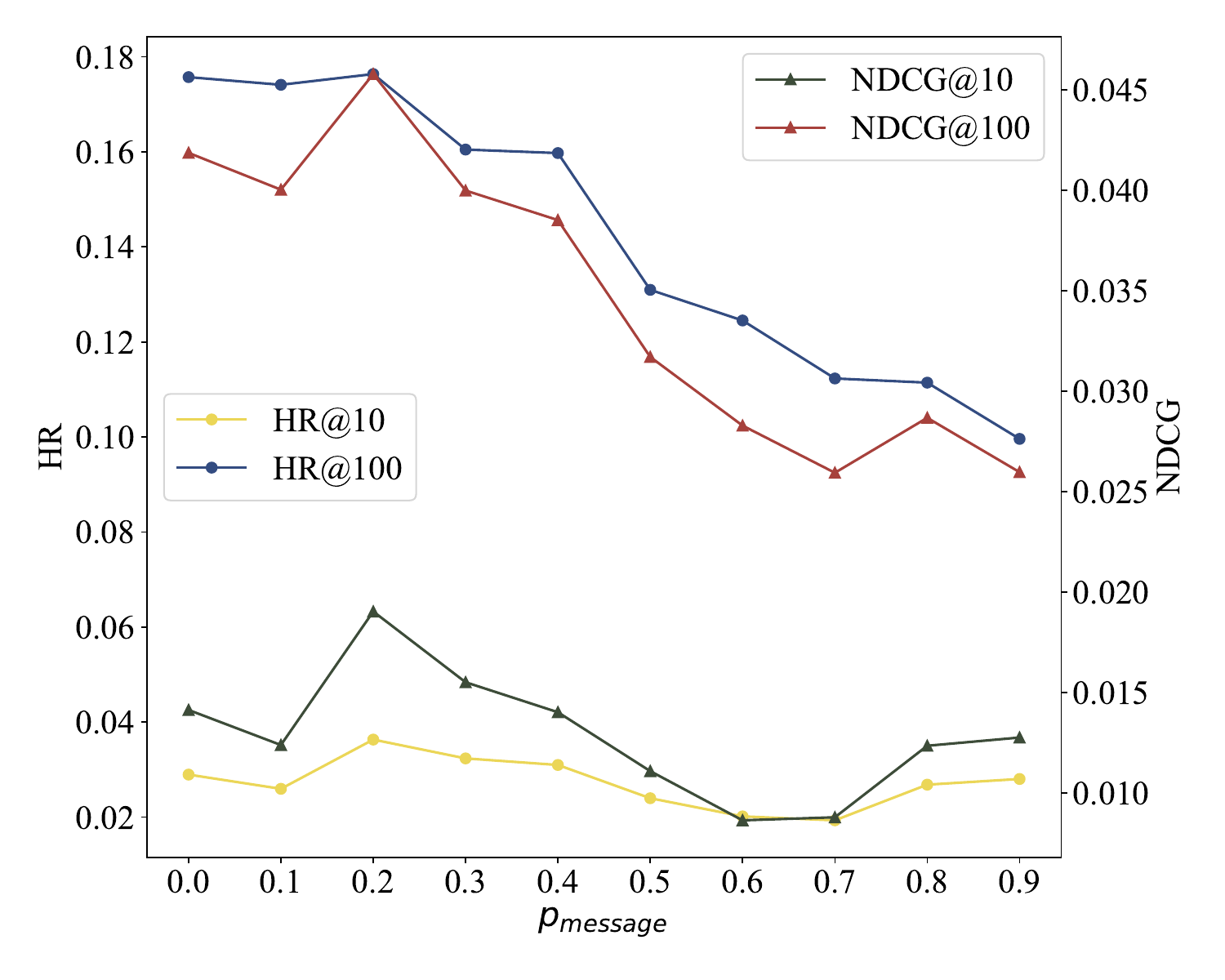}
    \caption{Visualization of results for different message dropout rates on Beibei.}
    \label{fig: dropout performance}
    \vspace{-0.5cm}
\end{figure}

\subsubsection{Impact of Synergy Alignment Task Weight}
We investigate the influence of the Synergy Alignment Task by analyzing its corresponding weight, $\lambda_a$, during the SWGCN optimization process. Specifically, we evaluate a range of values $\lambda_a \in \{0.1,0.2,0.3,0.4,$ $0.5,0.6,0.7,0.8,0.9\}$ using the Beibei and Taobao. The fluctuations in model performance are visualized in ~\ref{fig: lambda performance}. Observations from these results reveal that, for the Taobao dataset, the recommendation performance improves progressively with increasing $\lambda_a$, reaching its maximum at $\lambda_a=0.9$. In contrast, the Beibei dataset exhibits a fluctuating trend with modest variance, achieving optimal performance at $\lambda_a=0.2$. This observation, in conjunction with the previous analysis of interaction noise for both datasets, suggests a potential hypothesis: datasets exhibiting higher levels of noise may benefit from a greater emphasis on the Synergy Alignment Task, while for datasets with lower noise levels, the influence of this task on model performance appears to be minimal. This hypothesis warrants further empirical investigation.

\subsubsection{Impact of Message Dropout Rate}
We analyze the sensitivity of SWGCN to the message dropout ratio by varying $p_{message} \in \{0.0, 0.1, \dots, 0.9\}$ on the Beibei dataset. According to the outcomes presented in ~\ref{fig: dropout performance}, the growth of $p_{message}$ leads to the recommendation performance of SWGCN rising before experiencing a decline, with the optimal performance achieved at a dropout ratio of 0.2. Furthermore, the coarse-grained recommendation performance of SWGCN demonstrates heightened sensitivity to changes in $p_{message}$, resulting in a significant deterioration in the HR@100 and NDCG@100 metrics.

\begin{figure*}[pos=!t]
    \centering
    \subfigure[User \#11389]{
        \includegraphics[width=0.27\linewidth]{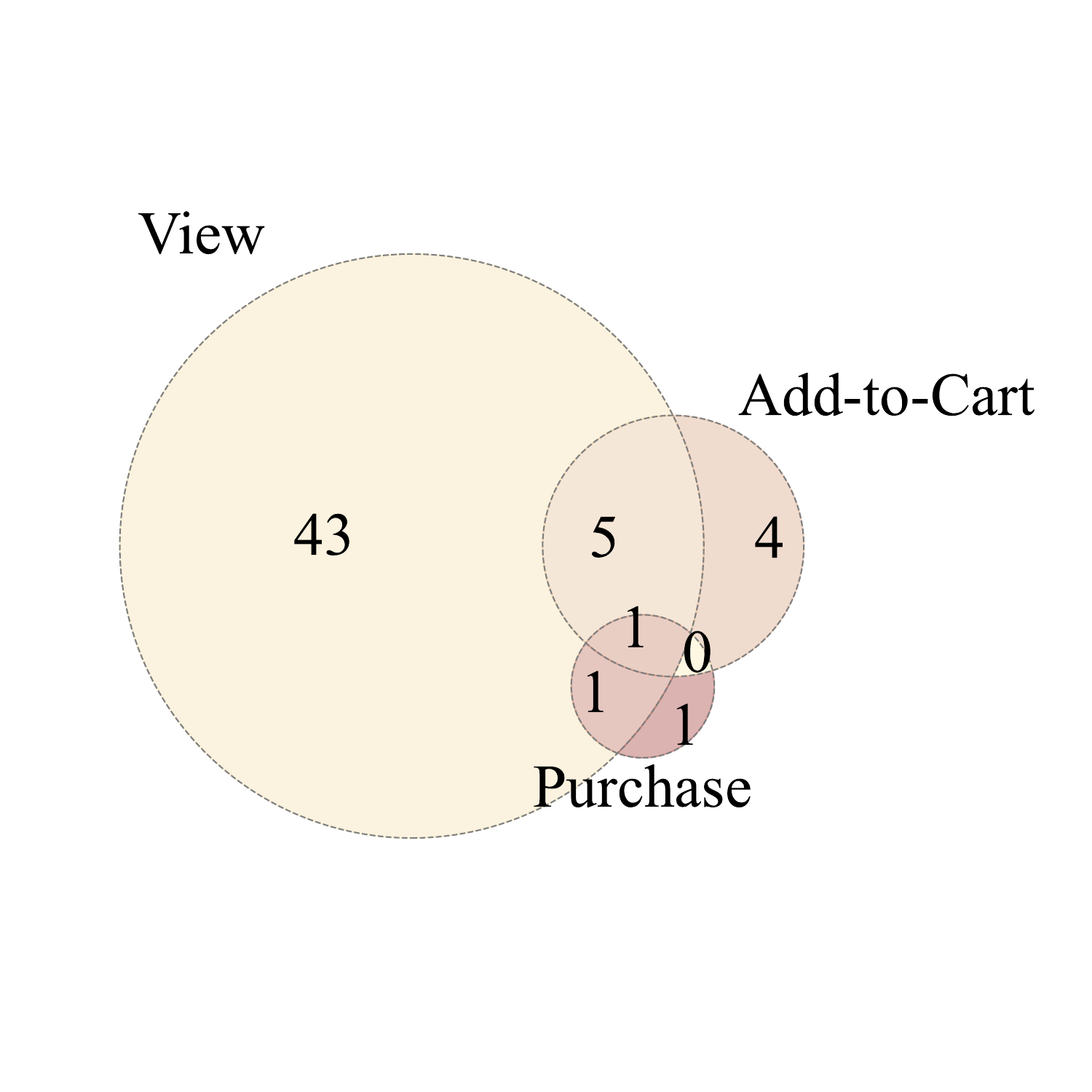}
    }
    \subfigure[User \#39258]{
	\includegraphics[width=0.27\linewidth]{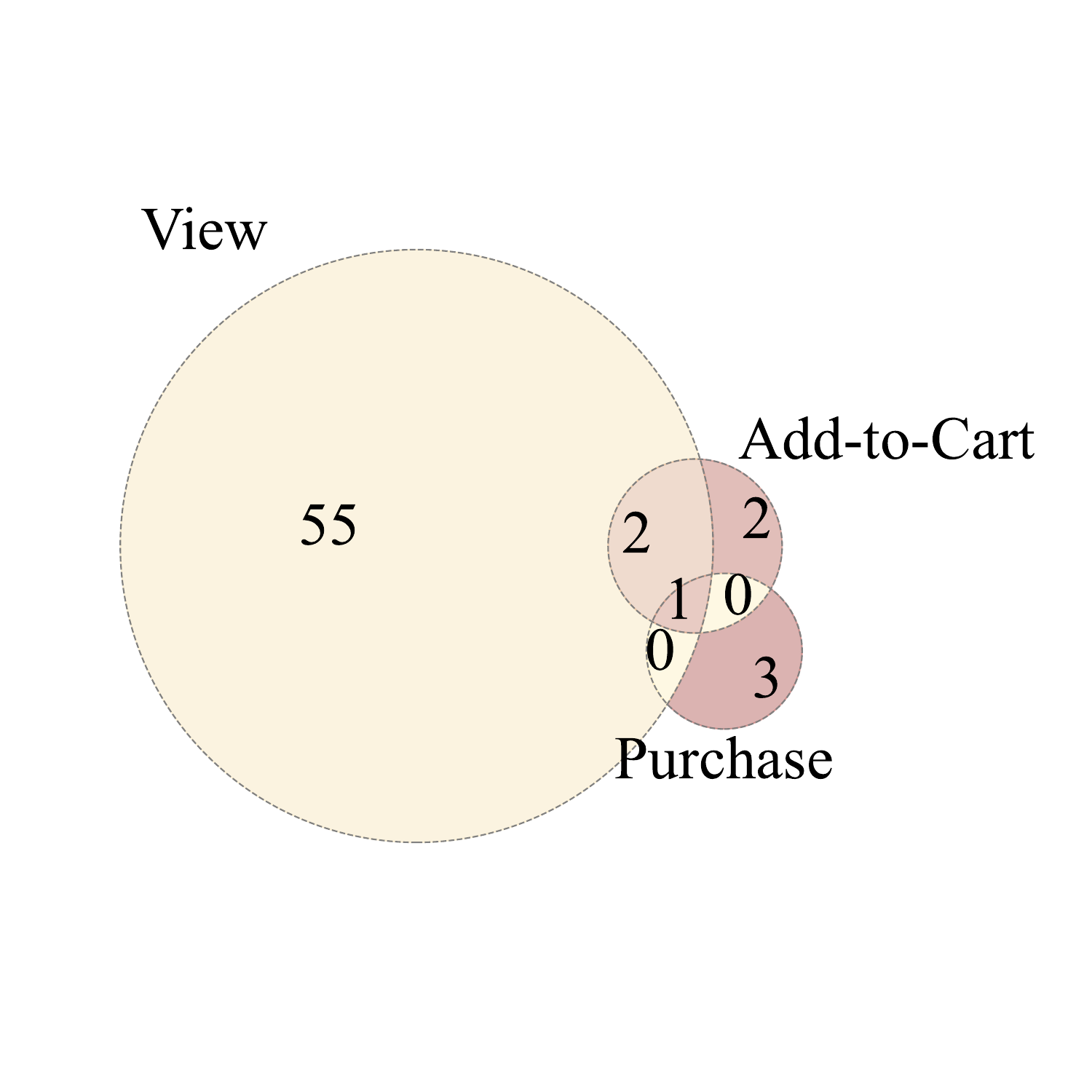}
    }
    \subfigure[User \#76748]{
	\includegraphics[width=0.27\linewidth]{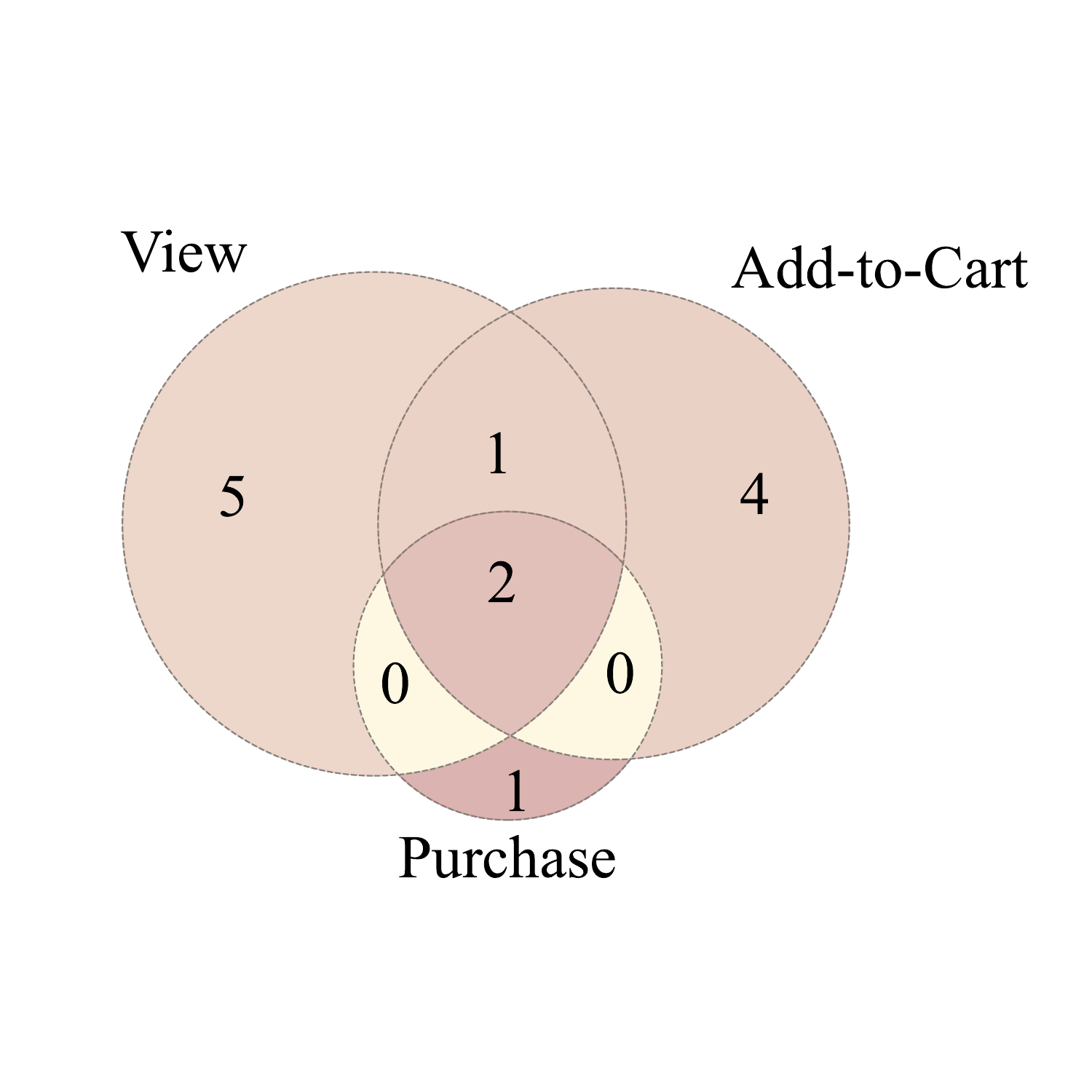}
    }
    \quad    
    \subfigure[User \#93932]{
    	\includegraphics[width=0.27\linewidth]{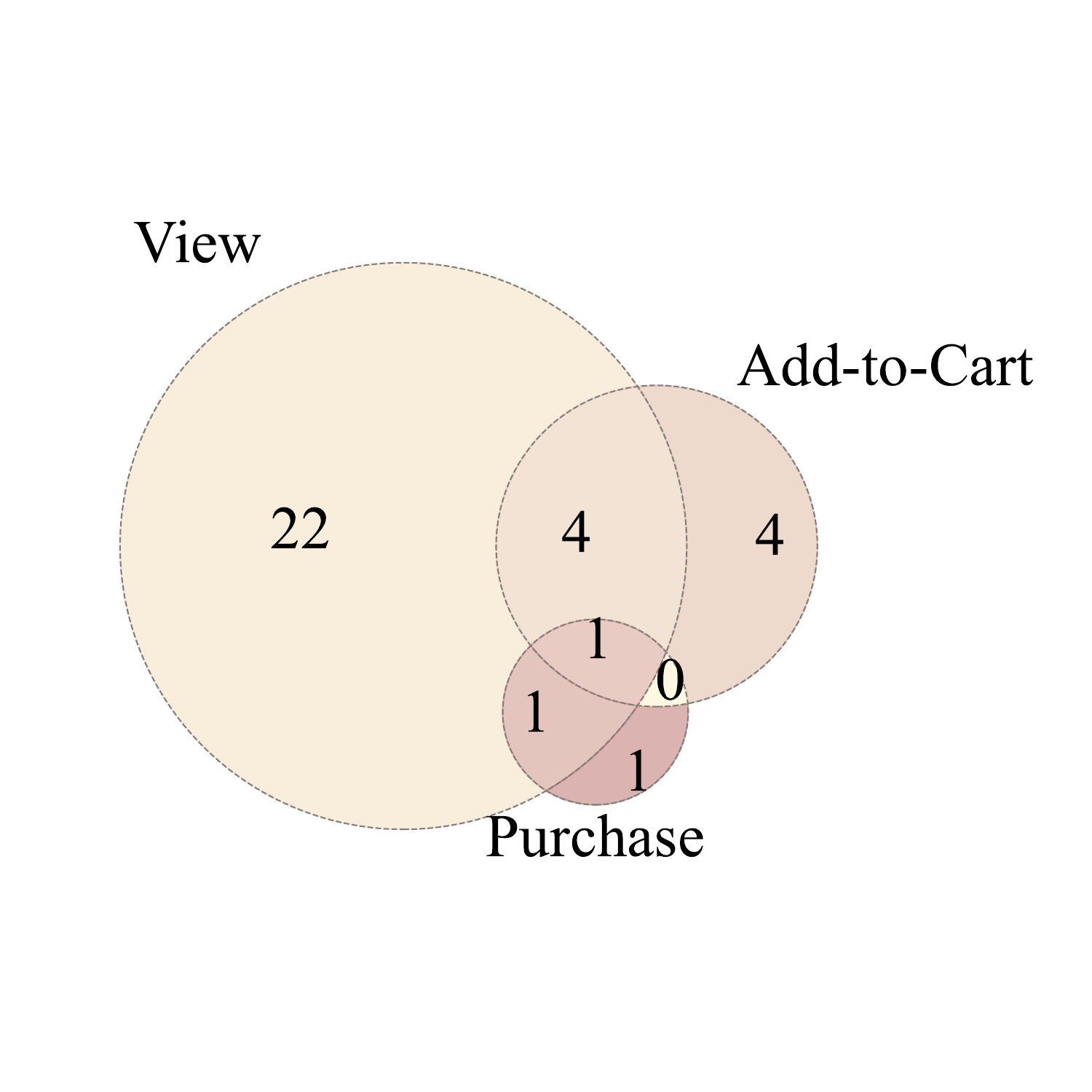}
    }
    \subfigure[User \#94575]{
	\includegraphics[width=0.27\linewidth]{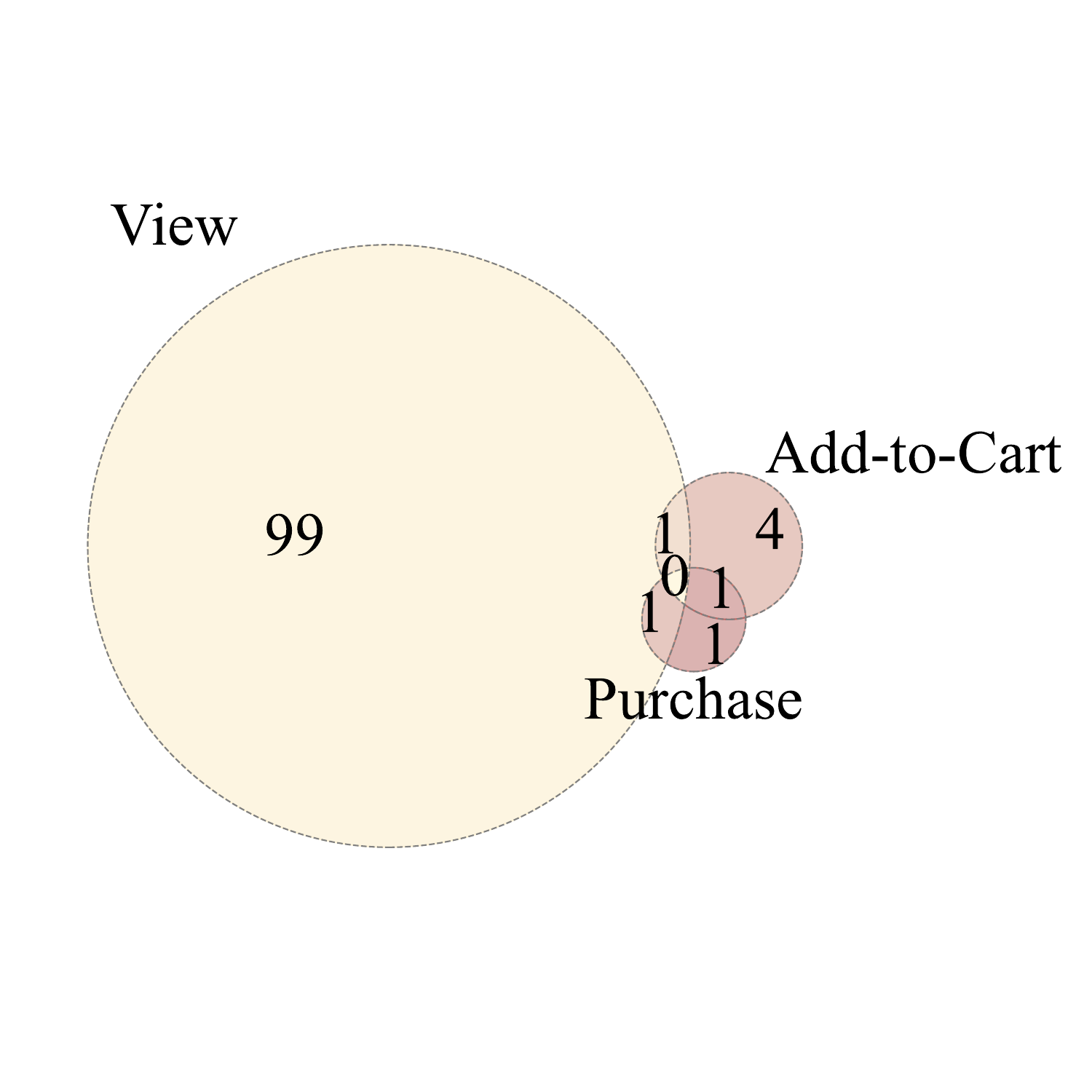}
    }
    \subfigure[User \#106291]{
	\includegraphics[width=0.27\linewidth]{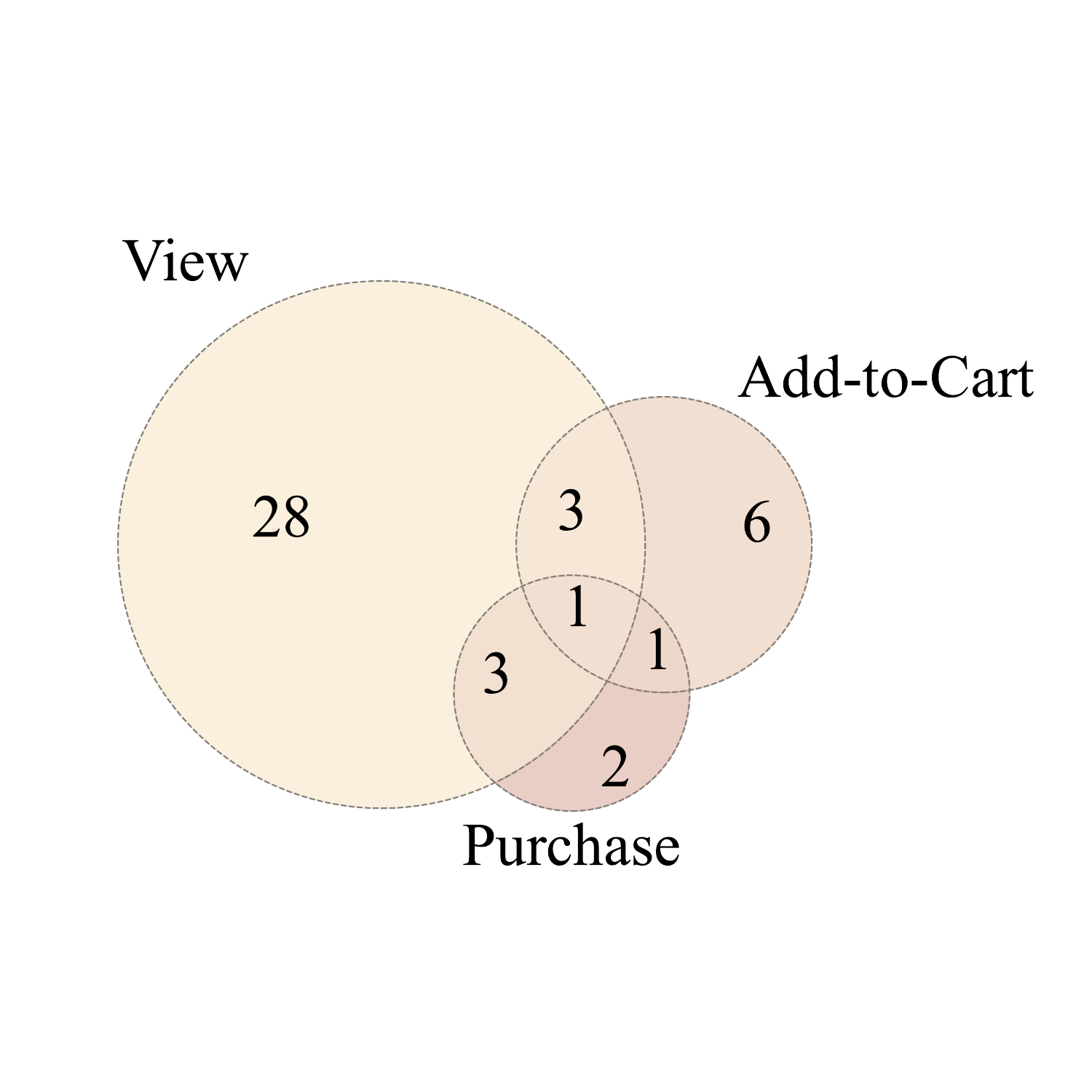}
    }
    \quad    
    \subfigure{
	\includegraphics[width=\linewidth]{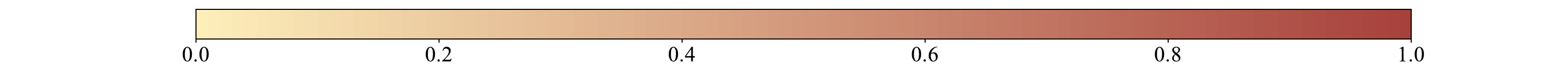}
    }
    \caption{Visualization of case study results.}
    \label{fig: case study}
    \vspace{-0.5cm}
\end{figure*}

\subsection{Case Study of User Synergy Interactions}
A case study is carried out to thoroughly assess the efficacy of SWGCN in enhancing recommendation performance by utilizing synergistic signals derived from auxiliary and target behaviors. Specifically, we extract the interaction histories of six users from the Taobao dataset, covering three distinct activity types: \textit{view}, \textit{add to cart}, and \textit{purchase}. Then, we draw Venn diagrams to partition the item sets associated with these behaviors. Furthermore, we calculate the average interaction weights for each subset of the Venn diagram and utilize heat maps to visualize these weights, with darker colors representing higher interaction weights. The findings of this investigation are illustrated in~\ref{fig: case study}. 

The analysis yields several significant insights. Firstly, it is evident that the weight attributed to the target behavioral interaction is predominant among all six users. This observation is consistent with the demands of real-world recommendation systems, which prioritize predicting target behavioral interactions. Thus, this finding substantiates the efficacy of SWGCN in discerning the importance of various behaviors. Secondly, items characterized by both auxiliary and target behavioral interactions demonstrate heightened significance, as illustrated by users \#11389, \#93932, and \#94575. This result not only underscores the ability of SWGCN to differentiate the importance of items within the same behavioral category but also indicates that SWGCN successfully captures the synergistic signals between auxiliary and target behaviors. Lastly, it is noted that items exhibiting three behavioral interactions do not necessarily possess greater significance. Specifically, the weights associated with the overlapping components of the three behaviors do not surpass those of items with two behavioral interactions, as evidenced by users \#11389, \#39258, \#93932, and \#106291. Intuitively, the presence of three behavioral interactions should imply an increased strength; however, the absence of enhanced interaction weights suggests a limitation within SWGCN to capture synergistic signals among three or more behaviors. Currently, SWGCN appears to support synergistic signals exclusively between the auxiliary behaviors and the target behavior.

\section{Discussion}

This section presents an in-depth evaluation of our empirical observations. We first elaborate on the significance of primary results and assess the relative importance of model components. Subsequently, we examine the model's stability across diverse hyperparameter configurations and discuss its feasibility for real-world implementation. Finally, we acknowledge existing constraints and identify prospective avenues for subsequent investigations.

\subsection{Remarks on Main Results}

The comparative experimental results across three real-world datasets offer several key insights into the behavior and effectiveness of SWGCN under varying conditions. First, the remarkable performance gain on the Taobao dataset—over 100\% improvement in both HR and NDCG—suggests that SWGCN can fully exploit dense synergistic patterns among behaviors when such signals are abundant and diverse. The prominent role of Target Preference Weigher and Synergy Alignment Task in this context highlights their ability to accurately differentiate and align multi-behavior preferences, especially when the co-interaction space is sufficiently large.

Second, the marginal improvements observed within the IJCAI and Beibei benchmarks suggest that the efficacy of our approach is contingent upon the intrinsic properties of the data. Specifically, the smaller performance margins on these datasets indicate that when auxiliary and target behaviors are less complementary or exhibit weaker co-occurrence structures, the utility of the Synergy Alignment Task becomes less critical. This is further validated by the SWGCN-T variant, where the absence of synergy between the auxiliary behaviors and the target behavior within Beibei dataset leads to performance fluctuations, emphasizing that synergy modeling is more effective when behavior overlap is substantial.

Third, it is observed that certain multi-behavior frameworks exhibit inferior efficacy compared to models based on a single interaction type (e.g., NGCF-MB~\cite{ngcf}), underscoring a robustness issue in current multi-behavior models. This suggests that not all multi-behavior architectures are equally generalizable, and naive fusion may introduce noise or lead to negative transfer if not properly weighted or aligned—an issue directly addressed by our weighted and synergy-aware design.

Lastly, the significant variance in performance gain across datasets raises an important point: the effectiveness of SWGCN hinges not only on model design but also on behavior distribution and interaction density. These observations suggest future directions in adaptive behavior selection, dynamic synergy modeling, and behavior-aware data augmentation.

\subsection{Relative Importance of Key Components}

We further reflect on the relative contribution of the two proposed modules—Target Preference Weigher and Synergy Alignment Task—to the overall performance of SWGCN.

As demonstrated in the ablation study (~\ref{fig: ablation study}), both components play indispensable roles in enhancing the recommendation quality. However, their relative impact differs across datasets and evaluation metrics. In particular, the Target Preference Weigher proves more vital to the model's success, as its elimination results in a sharper deterioration of results than the omission of the Synergy Alignment Task. For instance, in the Taobao dataset, the ``w/o TPW'' variant shows a dramatic drop of 38.53\% in HR and 48.23\% in NDCG, whereas the ``w/o $L_{SAT}$'' variant results in a relatively lower decline of 24.70\% and 32.79\% in the respective metrics. A similar trend is observed in the Beibei dataset.

These results suggest that the Target Preference Weigher is the more dominant factor in determining recommendation accuracy, as it directly governs the modeling of interaction importance within the target behavior. Nevertheless, the Synergy Alignment Task also proves crucial, particularly for datasets with richer auxiliary behavior signals, as it enhances the alignment between auxiliary and target behavior preferences.

Overall, we argue that the two components are complementary: the Target Preference Weigher primarily strengthens intra-behavior preference modeling, while the Synergy Alignment Task bridges multi-behavior signals. Their combined scheme contributes to the consistent efficacy and cross-domain utility of SWGCN among multiple datasets.

\subsection{Hyperparameter Robustness and Practical Deployment}

The sensitivity analysis of key hyperparameters in Section~\ref{self_loop_weight_exp} provides critical insights into the robustness and practical deployment of SWGCN. We observe that although the performance of SWGCN is influenced by parameters such as self-loop weight $\lambda_s$, Synergy Alignment Task weight $\lambda_a$, and message dropout rate $p_{message}$, the overall performance remains relatively stable within a reasonable range of each parameter. This suggests a degree of robustness and tolerance to non-optimal settings, which is particularly valuable for real-world applications where exhaustive hyperparameter tuning may be infeasible.

For instance, in the Beibei dataset, the optimal $\lambda_s$ is found at 0.2, closely aligning with the dominant interaction weight range of the target behavior. Meanwhile, the performance curve exhibits only modest fluctuations across the broader range of $\lambda_s \in [0.1, 0.4]$, indicating that SWGCN can adapt well to different graph characteristics with minimal tuning. In a similar vein, the alignment coefficient $\lambda_a$ exerts a greater effect on the Taobao benchmark compared with Beibei, likely due to Taobao's higher noise levels. This reinforces the notion that our model can selectively bolster alignment learning to counteract interference. This adaptability enhances the method's generalization to heterogeneous behavior scenarios.

The analysis of $p_{message}$ further reinforces this conclusion. While message dropout introduces stochasticity during training, we find that moderate dropout rates (e.g., 0.2) can improve generalization by preventing overfitting to local interaction structures. However, exceeding this threshold degrades performance, especially in top-$K$ ranking metrics, which indicates the necessity of focused rather than exhaustive tuning.

In practical deployment, we recommend initializing $\lambda_s$, $\lambda_a$, and $p_{message}$ using the empirically observed optimal ranges (e.g., $\lambda_s \in [0.2, 1.0]$, $\lambda_a \in [0.2, 0.9]$, $p_{message} \in [0.1, 0.3]$) and performing lightweight grid search or adaptive tuning if necessary. Overall, these findings highlight that SWGCN balances flexibility with stability, making it suitable for deployment in diverse and noisy real-world multi-behavior recommendation environments.

\subsection{Limitations and Future Work}

While SWGCN demonstrates superior performance and strong generalizability across four behavior recommendation tasks, several limitations remain that warrant further exploration. First, our current design primarily focuses on modeling synergistic signals among up to four behaviors. As revealed in our case study, the model exhibits reduced effectiveness in capturing complex synergy when involving three or more behaviors simultaneously. This limitation suggests the need for more advanced strategies capable of modeling high-order behavior interactions.

Second, the scope of our evaluation is confined to publicly available and relatively well-structured e-commerce datasets. In practice, real-world platforms often involve noisier, incomplete, or more heterogeneous data, and may contain additional behavior types such as consultations, clickstream sequences, or multi-modal interactions (e.g., image views, voice queries). These richer but messier data settings pose new challenges for modeling and generalization.

Future work may consider extending SWGCN to a more flexible architecture that can dynamically accommodate an arbitrary number of behaviors with varying semantics and data quality. Additionally, incorporating heterogeneous or multi-modal signals from diverse platforms could further improve the practical applicability of the model. Exploring these directions will not only enhance model robustness but also bridge the gap between academic benchmarks and real-world deployment.

\section{Conclusion}

In this research, we focus on tackling the challenges of synergistic interactions and internal heterogeneity that are often neglected by conventional multi-interaction models. In response to these limitations, we introduce a distinctive recommendation paradigm, denoted as \textbf{S}ynergy \textbf{W}eighted \textbf{G}raph \textbf{C}onvolutional \textbf{N}etwork (\textbf{SWGCN}). The framework incorporates a Target Preference Weigher module and a Synergy Alignment Task to capture fine-grained interaction intensities and inter-behavior dependencies. By refining user-item interaction weights, our framework significantly enhances the representation learning of graph neural networks. Thorough experimentation on three openly available real-world datasets indicates that SWGCN surpasses current benchmarks in performance on all datasets examined, affirming its superiority and generalizability. However, a case study reveals a limitation in SWGCN in modeling synergistic signals among three or more behaviors. Consequently, there is a compelling need for advanced strategies to navigate the intricate hurdles found in recommendation systems involving multiple interactions.

\section{Acknowledgements}
This work was supported by the National Natural Science Foundation of China (No.62376043, No.62002035), and the Natural Science Foundation of Chongqing (No.CSTB2023NSCQ-MSX0091).

\section{Declaration of generative AI and AI-assisted technologies in the writing process}
In the process of preparing this manuscript, the author leveraged ChatGLM to assist with stylistic and grammatical improvements. All AI-facilitated suggestions were carefully evaluated and refined by the author, who holds full responsibility for the published findings and text.

\printcredits

\bibliographystyle{elsarticle-num}

\bibliography{refs-r1}

\end{document}